\documentclass[journal]{IEEEtran}

\usepackage{times}
\usepackage{mathrsfs}
\usepackage{epsfig}
\usepackage{graphicx}
\usepackage{amsmath}
\usepackage{amssymb}
\usepackage{graphicx}
\usepackage{cite}
\usepackage{enumerate}
\usepackage{cases}
\usepackage{multirow}
\usepackage{verbatim}
\usepackage{amssymb}
\usepackage{CJK}
\usepackage{algorithm}
\usepackage{algorithmicx}
\usepackage{algpseudocode}
\usepackage{color}
\usepackage{bm}
\usepackage{booktabs}
\usepackage[colorlinks,linkcolor=red]{hyperref}
\usepackage{amssymb}
\usepackage{makecell}

\newcommand{\etal}{\textit{et al}.}
\newcommand{\ie}{\textit{i}.\textit{e}.}
\newcommand{\eg}{\textit{e}.\textit{g}.}
\newcommand{\etc}{\textit{etc}}

\begin{document}

\title{BCS-Net: Boundary, Context and Semantic for Automatic COVID-19 Lung Infection Segmentation from CT Images}

\author
{
Runmin Cong,~\IEEEmembership{Member,~IEEE,} Haowei Yang, Qiuping Jiang, Wei Gao, Haisheng Li, Cong Wang,\\ Yao Zhao,~\IEEEmembership{Senior Member,~IEEE,} and Sam Kwong,~\IEEEmembership{Fellow,~IEEE}

\thanks{Runmin Cong is with the Institute of Information Science, Beijing Jiaotong University, Beijing 100044, China, also with Beijing Key Laboratory of Big Data Technology for Food Safety, Beijing Technology and Business University, Beijing 100048, China, and also with the Department of Computer Science, City University of Hong Kong, Hong Kong SAR, China (e-mail: rmcong@bjtu.edu.cn).}
\thanks{Haowei Yang and Yao Zhao are with the Institute of Information Science, Beijing Jiaotong University, Beijing 100044, China, and also with the Beijing Key Laboratory of Advanced Information Science and Network Technology, Beijing 100044, China (e-mail: hwyang@bjtu.edu.cn; yzhao@bjtu.edu.cn).}
\thanks{Qiuping Jiang is with the School of Information Science and Engineering, Ningbo University, Ningbo 315211, China (e-mail: jiangqiuping@nbu.edu.cn).}
\thanks{Wei Gao is with the School of Electronic and Computer Engineering, Peking University Shenzhen Graduate School, Shenzhen 518055, China, and also with Peng Cheng Laboratory, Shenzhen 518055, China (e-mail: gaowei262@pku.edu.cn).}
\thanks{Haisheng Li is with the Beijing Key Laboratory of Big Data Technology for Food Safety, Beijing Technology and Business University, Beijing 100048, China (e-mail: li\_haisheng@163.com).}
\thanks{Cong Wang is with the Distributed and Parallel Software Lab, Huawei Technologies, Shenzhen 518129, China (e-mail: wangcong64@huawei.com).}
\thanks{Sam Kwong is with the Department of Computer Science, City University of Hong Kong, Hong Kong SAR, China, and also with the City University of Hong Kong Shenzhen Research Institute, Shenzhen 51800, China (e-mail: cssamk@cityu.edu.hk).}
}

\markboth{IEEE Transactions on Instrumentation and Measurement}
{Shell \MakeLowercase{\textit{et al.}}: Bare Demo of IEEEtran.cls for IEEE Journals}
\maketitle

\begin{abstract}
The spread of COVID-19 has brought a huge disaster to the world, and the automatic segmentation of infection regions can help doctors to make diagnosis quickly and reduce workload. However, there are several challenges for the accurate and complete segmentation, such as the scattered infection area distribution, complex background noises, and blurred segmentation boundaries. To this end, in this paper, we propose a novel network for automatic COVID-19 lung infection segmentation from CT images, named BCS-Net, which considers the boundary, context, and semantic attributes. The BCS-Net follows an encoder-decoder architecture, and more designs focus on the decoder stage that includes three progressively Boundary-Context-Semantic Reconstruction (BCSR) blocks. In each BCSR block, the attention-guided global context (AGGC) module is designed to learn the most valuable encoder features for decoder by highlighting the important spatial and boundary locations and modeling the global context dependence. Besides, a semantic guidance (SG) unit generates the semantic guidance map to refine the decoder features by aggregating multi-scale high-level features at the intermediate resolution. Extensive experiments demonstrate that our proposed framework outperforms the existing competitors both qualitatively and quantitatively.
\end{abstract}


\IEEEpeerreviewmaketitle

\section{Introduction}
\IEEEPARstart{G}{lobally}, as of March 2022, more than 452 million confirmed cases of COVID-19 have been reported to WHO, including more than 6 million deaths.
Especially, the new wave of epidemics caused by the Delta and Omicron variants of COVID that broke out in India, South Korea, and Hong Kong from 2021 is more contagious, the global epidemic situation still cannot be relaxed.

\begin{figure}[!t]
\centerline{\includegraphics[width=\columnwidth]{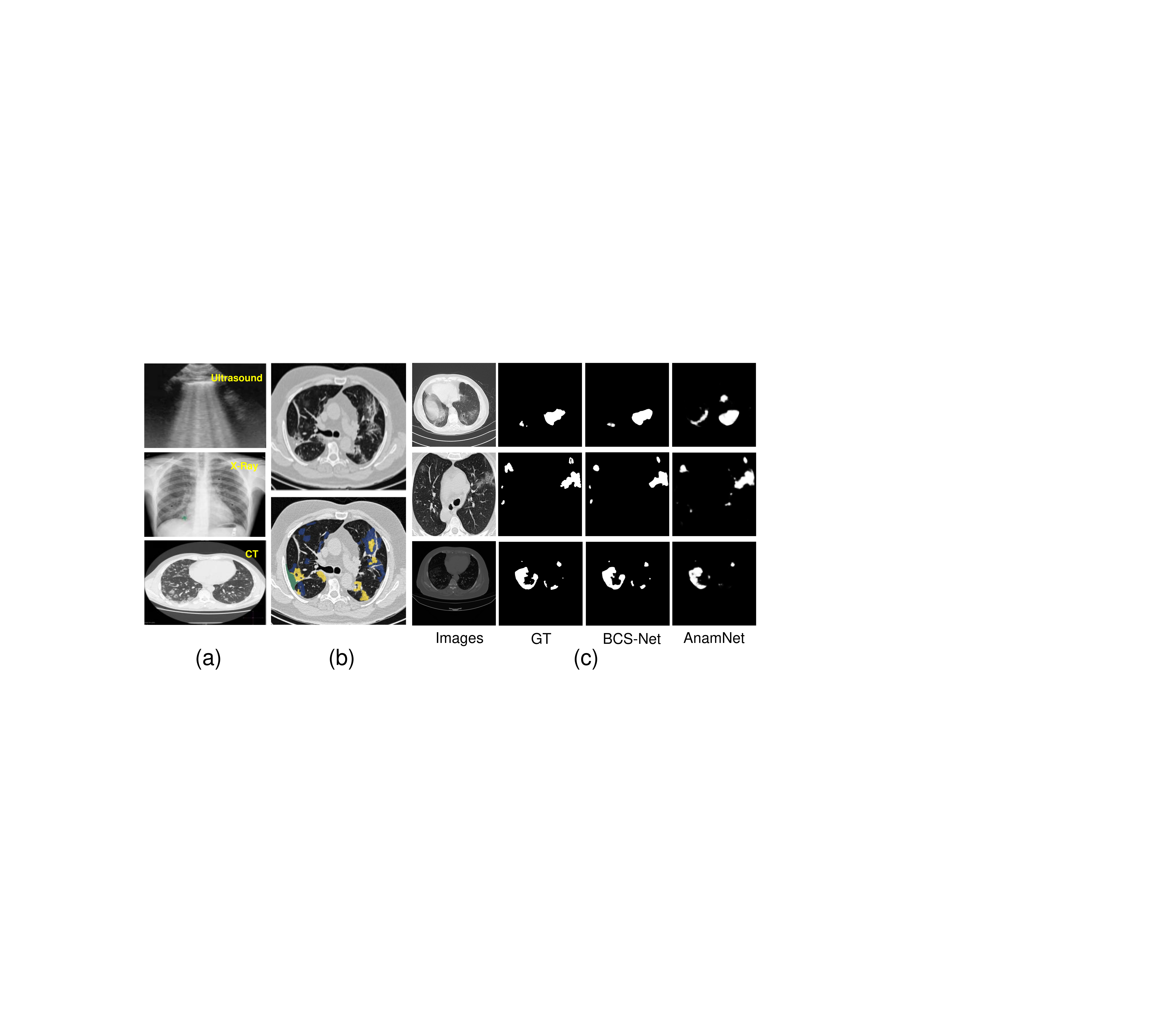}}
\caption{(a) Some images taken in the lung with three different imaging devices. (b) The COVID-19 infected lung CT images, where the ground-glass opacity, consolidation and pleural effusion are marked in blue, yellow and green respectively. (c) Visual examples of different COVID-19 infection segmentation methods.}
\label{fig1}
\end{figure}

As a rapid and large-scale COVID-19 testing method, the RT-PCR testing has been widely adopted worldwide, but its false negative rate is as high as 17\% to 25.5\%, which is only suitable for preliminary screening.
In clinical practice, to make a definite diagnosis of suspected cases and determine an appropriate treatment plan, lung imaging interpretation by ultrasound \cite{boundariesatten,Ultrasound}, X-rays \cite{r6,r9,r10} or computed tomography (CT) \cite{r17,wang2020weakly,infnet,AnamNet,COPLENet,r20,DBLP:journals/tim/Roy22,2022SSA} is an indispensable link.
However, different imaging devices have their own advantages. Ultrasound can propagate in a certain direction and penetrate objects. Based on the principle that ultrasonic waves generate echoes, we can collect and display such echoes on the screen through instruments to understand the internal structure of objects, as shown in the first row of Fig. \ref{fig1}(a). X-ray examination uses the penetrating power of X-rays that travel through the body to a detector on the other side. Due to the different absorption of X-rays by bones, muscle tissues and air, an image with different gray levels from black to white is formed, as shown in the second row of Fig. \ref{fig1}(a). Computed tomography (CT) scans a certain part of the human body with X-ray beam to obtain a cross-sectional or three-dimensional image of the part being examined, which can clearly display the organs and structures, as shown in the third row of Fig. \ref{fig1}(a) and  Fig. \ref{fig1}(b). Comparing these three imaging techniques, ultrasound can detect the consolidation of lung tissue, and X-ray has a good ability to show the presence of large areas of infected areas. However, these are based on the fact that the virus infected by the patient has developed to a certain stage. By contrast, the chest CT scans show clearer structures, have better sensitivity and accuracy than traditional chest X-rays and ultrasound waves, and can detect subtle lesions in the early stages of lung infection. Based on the characteristics of different data, ultrasound data and X-ray data are more suitable for follow-up observation of patients with pulmonary infection after diagnosis, and their convenience and low cost have great advantages. However, CT images show the irreplaceable role of the lung cavity in the definite diagnosis and early diagnosis, which has been accepted and adopted by the international medical community.
%The spread of COVID-19 is very fast and highly sporadic, and it is likely to cause a surge in patients in the short term, which undoubtedly puts enormous pressure on the medical system. In order to make a definite diagnosis of suspected cases and determine an appropriate treatment plan, doctors usually need to screen the lung CT images of patients.
However, problems such as the surge of patients, the shortage of professional doctors, and the huge workload can easily delay diagnosis and even cause misjudgment. To this end, artificial intelligence technology can be used to automatically interpret the CT images, and provide auxiliary diagnosis information such as infection area classification and segmentation, thereby reducing the workload of doctors and improving the accuracy of screening. From this point, correct segmentation of the COVID-19 infection regions is critical for timely treatment of patients, and the research work in this paper takes this as a starting point to achieve accurate and efficient COVID-19 lung infection segmentation. Therefore, in the research sense, our proposed method in this paper uses artificial intelligence technology for COVID-19 infection segmentation, which can improve the segmentation accuracy and reduce the work pressure of radiologists.
%To this end, in this paper, we focus on the automatic COVID-19 lung infection segmentation from CT images, which belongs to the field of medical image segmentation.

% Moreover, based on the lung image, the confirmed cases can be classified to provide different treatments.
% X-rays are three-dimensional volume overlapped on the two-dimensional plane, so only the projection can be seen, making it easy to cover up the early small lesion regions for the projection image.
% By contrast, CT screening is a tomographic imaging of the target area, just like a scanner, which can scan the whole lung space and clearly exhibit the fine lesions.
% According to the imaging manifestations of the lung CT image, we can determine the development of the patient's condition,
% such as multiple small patchy shadows and interstitial changes of the lung, multiple ground-glass opacity (GGO) manifestations, and infiltrating shadows of both lungs. Therefore, the rapid and accurate interpretation of the lung infection area can provide an auxiliary role for the doctor's diagnosis.
% To this end, in this paper, we focus on the automatic COVID-19 lung infection segmentation from CT images, which belongs to the field of medical image segmentation.

In recent years, the vigorous development of deep learning has greatly promoted the development of computer vision-related fields \cite{crm2020tc,crmbridgenet,crmunderwater,crmDPANet,crmglnet,crmRRNet,crm2022rsi,crmDBLP:journals/spl/HuJCGS21,crmDBLP:journals/tip/LiAHCGR21,crmDBLP:journals/tip/WenYZCSZZBD21,crmDBLP:journals/tmm/MaoJCGSK22,crmCoADNet,crmACMMM20-1,crm-nc,crm2019tgrs,crm-acmmm,crmijcai20,crmgcl2019tip,crmGCPANet}. Among them, medical image segmentation algorithms have made great progress and achieved a qualitative leap in performance, such as lung nodules segmentation \cite{intro-4wang2017central}, breast-region segmentation \cite{DBLP:journals/tim/PramanikGBN20}, brain and brain-tumor segmentation \cite{intro-5cherukuri2017learning},  Brain Image Synthesis \cite{crmbrain}, polyp segmentation \cite{crmpolyp}, COVID-19 infection detection \cite{DBLP:journals/tim/DairiHS22}, COVID-19 forecasting \cite{DBLP:journals/tim/SharmaKMR21}, and COVID-19 infection segmentation \cite{DBLP:journals/tim/Roy22,infnet}  \etc. However, differences in imaging equipment and disease characteristics make it difficult to use a unified segmentation model for different diseases.
Researchers often need to design some unique modules to better achieve lesion area segmentation. Observing an example shown in Fig. \ref{fig1}, we can summarize three difficult problems that the COVID-19 lung infection segmentation models need to solve:
\begin{enumerate}[(1)]
\item The infection regions of COVID-19 are very scattered, with many isolated areas of various sizes, which are very challenging for complete detection. Moreover, the sizes of the infected areas varies greatly, which makes it more difficult to accurately detect the infected areas of different scales. In addition, the boundary of the infection region is not easy to segment accurately and sharply. To this end, we can seek a solution from the perspective of encoder features containing relatively rich and effective information, which can be used to guide the feature learning in the decoder stage. Concretely, we propose an Attention-Guided Global Context (AGGC) module to select the most valuable information from encoder features for decoder, where the spatial attention and boundary attention are used to provide more accurate important spatial and boundary guidance information, and the global context modeling unit is used to model the global context dependence and constrain the generation of more complete segmentation result.
% The infected areas are relatively scattered, which poses a challenge to the segmentation accuracy and completeness.
% As can be seen, the infected areas are usually not concentrated, but scattered in different positions of the lung, which makes it easy to cause missed detection of the object or part of the object.
% Moreover, the sizes of the infected areas varies greatly, which makes it more difficult to accurately detect the infected areas of different scales.
% Faced with these problems, we can seek a solution from the perspective of encoder features containing relatively rich and effective information, which can be used to guide the feature learning in the decoder stage.
% To this end, we propose an Attention-Guided Global Context (AGGC) module in this paper to select the most valuable information from encoder features for decoder.
%The AGGC module not only highlights the important spatial and boundary locations to provide more accurate positioning guidance information, but also models the global context dependence to constrain the generation of more complete segmentation result.

\item The infected areas have many detailed boundaries, and more attention needs to be paid to the clearness and sharpness of the boundaries in the segmentation results.
From a clinical point of view, the boundary of the infected area is of great significance for diagnosis. However, due to the pooling or up-sampling/down-sampling operation, the CNN-based method may blur the boundaries of the segmentation result. To generate a clearer and sharper boundary, we introduce boundary guidance in the proposed network.
In addition to introducing the supervised learning of the boundary map, we also treat the generated boundary map as an attention weight to highlight the encoder features in the AGGC module and allow the boundary constraint to play a greater role in the network.

\item For the COVID-19 infected lung CT image, although the processed data is only limited to the image of the lung area, the background regions are relatively complex and there are still a lot of interferences, such as inflamed areas that are not infected by COVID-19. Therefore, it is crucial to effectively suppress background noises. Considering that the high-level features including more semantic and category attributes can be used to suppress complex backgrounds, we design a Semantic Guidance (SG) unit to aggregate multi-scale high-level features and formulate a semantic guidance map to refine the decoder features. Moreover, in order to alleviate the information loss from multiple sampling operations and maintain the correlation between feature maps of adjacent scales, the multi-scale feature fusion is unified on the intermediate resolution to generate the semantic guidance map.
% The backgrounds in the lung CT images are relatively complex with more interferences, and thus it is necessary to effectively suppress the background noises.
% Although the processed data is only limited to the image of the lung area, there are still many interferences, such as inflamed areas that are not infected by COVID-19.
% For this issue, the high-level features including more semantic and category attributes can be used to suppress complex backgrounds.
% Thus, a Semantic Guidance (SG) unit is designed to aggregate multi-scale high-level features and formulate a semantic guidance map to refine the decoder features.
% Moreover, in order to alleviate the information loss from multiple sampling operations and maintain the correlation between feature maps of adjacent scales, the fusion of multi-scale high-level features are unified on the intermediate resolution to generate the semantic guidance map.
\end{enumerate}

In summary, we propose an automatic COVID-19 lung infection segmentation network with the encoder-decoder structure, equipped with three Boundary-Context-Semantic Reconstruction (BCSR) blocks. Although the encoder-decoder architecture is a common structure in medical image segmentation, we make a delicate design in it to cope with the special characteristics of the COVID-19 infection segmentation task, such as relatively scattered infected regions, more boundary details, complex backgrounds and noisy interferences. First, in order to address the problem of inaccurate and incomplete detection caused by the scattered infected regions, we design an AGGC module to select the most valuable information from encoder features for decoder, which highlights the important spatial and boundary locations in an attention manner, and models the global context dependence for the complete segmentation. Second, in order to achieve the sharp boundaries of the final segmentation result, we introduce the BA unit by adding the boundary supervised learning and boundary map refinement. Third, in order to suppress the background interferences, a SG unit is designed to aggregate multi-scale high-level features and formulate a semantic guidance map to refine the decoder features. All modules cooperate with each other to make our network achieve the competitive performance. As shown in Fig. \ref{fig1}(c), our proposed method has advantages in detection accuracy, completeness, clarity and sharpness.
The main contributions of this paper are as follows:
\begin{itemize}
\item An end-to-end network, named BCS-Net, is proposed to achieve automatic COVID-19 lung infection segmentation from CT images, which models the boundary constraint, context relationship, and semantic guidance. Moreover, our network achieves the competitive performance on the publicly available dataset both qualitatively and quantitatively.
\item An AGGC module is designed to filter the most valuable encoder information for decoder, which highlights the important spatial and boundary locations in an attention manner, and models the global context dependence for the complete segmentation.
\item A SG unit is proposed to aggregate multi-scale high-level features and generate the semantic guidance map to refine the decoder features, in which the multi-scale fusion is unified on the intermediate resolution to alleviate the information loss and maintain the correlation between adjacent scales.
\end{itemize}

\section{RELATED WORK}
\begin{figure*}[!t]
\centerline{\includegraphics[width=1\textwidth]{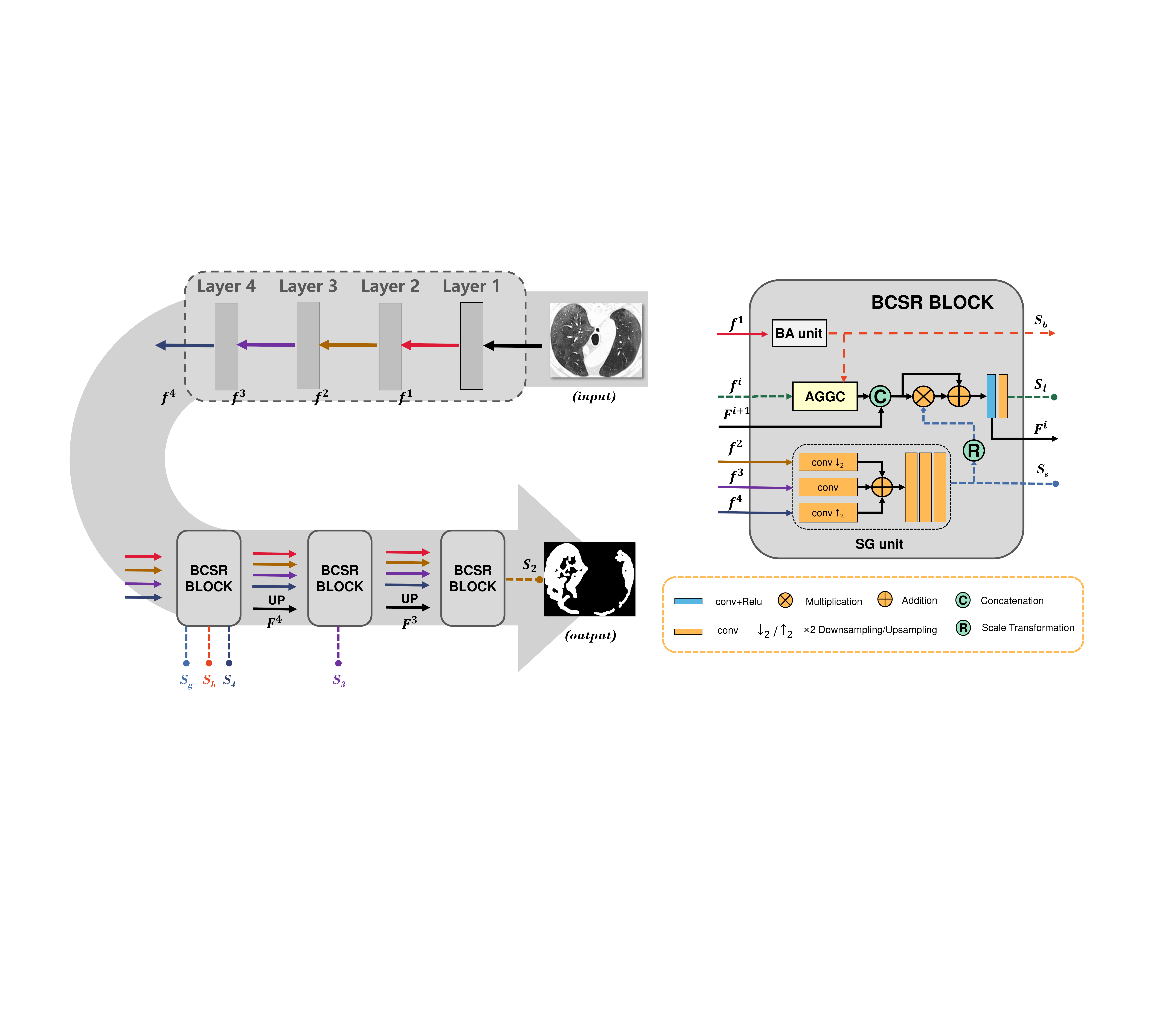}}
\caption{Illustration of the overall framework of proposed network. The input CT slice image is first embedded into the backbone with four layers to extract the multi-level features, then we utilize the stacked Boundary-Context-Semantic Reconstruction (BCSR) blocks to progressively reconstruct the segmentation results.
The BCSR block consists of Attention-Guided Global Context (AGGC) module, Semantic Guidance (SG) unit, Boundary Attention (BA) unit. The framework finally produces five prediction maps, in which the $S_2$ map is the final segmentation result.
}
\label{fig2}
\end{figure*}
%Since the outbreak of COVID-19, medical imaging provides an important supplement for real-time diagnosis and treatment, among which the auxiliary segmentation and prediction technology based on deep learning has played a significant role.
In this section, we will briefly introduce some related works in COVID-19 diagnosis based on deep learning for CXR and CT images, such as  classification, segmentation.
There exist lots of deep learning-based segmentation methods, such as Mask R-CNN \cite{maskrcnn}, YOLACT \cite{YOLACT}, Deeplab \cite{deeplab}, FCN  \cite{FCN}, \etc, which have made great achievements in segmentation tasks of various scenes and can be used for the medical segmentation task. But there are some differences between medical images and traditional ordinary images. Taking the COVID-19-infected lung CT image as an example, the areas to be segmented are usually scattered, with more details, complex backgrounds, and more interfering noises. Therefore, after the emergence of the U-Net \cite{unet}, it has gradually become the infrastructure for medical image segmentation and even general image segmentation tasks. Its success lies in the fact that a symmetric encoder-decoder structure with skip connections can better and more comprehensively utilize features at different levels, thereby generating more task-discriminative representations and improving performance.

% For the CXR datasets for COVID-19, there are few datasets available to the public \cite{r5}. In \cite{r6}, the problem of data shortage was alleviated by using a model pre-trained on ImageNet. In addition, transfer learning was introduced to achieve the COVID-19 classification and detection \cite{r9}.
% Oh \etal \cite{r4} proposed a patch-based network to overcome the shortage of CXR datasets, which can diagnosis COVID-19 with a relatively small number of datasets.
% In general, CXR is mostly used for COVID-19 classification task, such as distinguishing the healthy lungs, non-COVID-19 pneumonia-infected lungs, or COVID-19-infected lungs.

Due to the high sensitivity and clarity of CT images, most of the current diagnosis of COVID-19 is based on CT images, including segmentation \cite{r17,wang2020weakly,infnet,AnamNet,COPLENet,r20}.
In practice, segmenting the key area from chest CT can provide useful information for medical staffs to diagnose the COVID-19, including GGO (Ground Glass Opacity) and consolidation. So far, many COVID-19 lung infection segmentation methods from CT Images based on deep learning have been proposed, and promising performance has been obtained.
Zhou \etal \cite{r17} proposed an automated segmentation network by integrating the spatial and channel attention mechanisms.
Fan \etal \cite{infnet} designed the parallel partial decoder, reverse attention, and edge-attention to boost the segmentation accuracy, and also provided a semi-supervised framework to alleviate the shortage of labeled data.
Paluru \etal \cite{AnamNet} proposed an anamorphic depth embedding-based lightweight CNN to segment anomalies in COVID-19 chest CT images, which incorporates fully convolutional anamorphic depth blocks with depth-wise squeezing and stretching after the down-sampling and up-sampling operations.
Wang \etal \cite{COPLENet} proposed a noise-robust learning framework based on self-ensembling of CNNs.
Han \etal \cite{r20} proposed an attention-based deep 3D multiple instance learning (AD3D-MIL) to achieve the goal of accurate and interpretable screening of COVID-19 from chest CT images.

In these previous works, some ingenious and effective structures were designed, which achieved good performance and brought us a lot of inspiration. However, the existing work does not fully consider the following two points: (1) the loss of semantic information caused by oversampling in multi-layer feature fusion; (2) the correlation between contextual information and semantic information plays an important role in complete segmentation. For the first issue, we use the middle scale as the output criterion in the SG unit to generate the semantic guidance mask, which avoids excessive up- and down-sampling of a single layer in the multi-layer feature fusion. As for the second issue, we design an AGGC module to select the most valuable information from encoder features for decoder, which can not only highlight the important spatial and boundary locations, but also perceive the global correlation between different areas.

\section{PROPOSED METHOD}
\subsection{Overview}
Fig. \ref{fig2} illustrates the overall framework of our proposed BCS-Net to achieve the COVID-19 lung infection segmentation from CT images. Our network follows an encoder-decoder architecture in an end-to-end manner, in which the backbone extractor \cite{p1-gao2019res2net} in the encoder stage aims to extract the top-down multi-level features, and the decoder stage includes three progressively Boundary-Context-Semantic Reconstruction (BCSR) blocks to learn the segmentation-related features and generate the segmentation mask. In each BCSR block, we jointly consider the global semantic guidance, context dependence modeling, and boundary attention refinement to provide more sufficient supplementary information for feature decoding. Specifically, in order to deal with the problem of scattered lesion location and irregular shape, we design an attention-guided global context (AGGC) module, which selects the most valuable encoder propagation features by performing attention and context modeling on the encoder features of the corresponding decoder layer. As shown in Fig. \ref{fig3}, the corresponding encoder features and boundary attention map generated by the Boundary Attention (BA) unit are fed into the AGGC module. The encoder features are first refined by the spatial attention and boundary attention to highlight the important spatial locations and boundary details. Then, the Global Context Modeling (GCM) unit is used to correlate the different locations to consistently detect the lesion regions and generate the enhanced encoder features. With the guidance of the enhanced encoder features, the initial decoder features of the current level are generated by combining with the previous decoder features generated by the previous BCSR block. Furthermore, considering the importance of high-level semantic information for suppressing irrelevant background noises, we design a Semantic Guidance (SG) unit to aggregate multi-scale high-level features and formulate a semantic guidance map, which is further combined with the decoder features by means of residual connection to obtain the corresponding final decoder features:
\begin{equation}
F^i=\delta(conv(\bar{F}^i+\bar{F}^i\odot S_s))
\end{equation}
where $\bar{F}^i=concat(F^{i+1}, f^i_{aggc})$, $concat(\cdot)$ is channel-wise concatenation operation, $f^i_{aggc}$ are the output features generated by the AGGC module, $S_s$ denotes the semantic guidance map generated by the SG unit, $\odot$ represents element-wise multiplication, $\delta(\cdot)$ denotes the ReLU activation, and $conv(\cdot)$ is a customized convolutional block.
At each BCSR block, we use a convolutional layer with the kernel size of $1\times1$ to produce the corresponding segmentation map $S_i$:
\begin{equation}
S_i=\sigma(conv_{1\times 1}(F^i))
\end{equation}
where $conv_{1\times1}$ denotes a convolutional layer with the kernel size of $1\times1$, and $\sigma(\cdot)$ is the Sigmoid activation. Our network produces five side-output maps $(S_b,S_s,S_4,S_3,S_2)$, where the output $S_2$ of the last BCSR block is used as the final infection segmentation mask.
%In following subsections, we will introduce the details of BCSR block including the AGGC module, SG unit, BA unit, and loss function.

\subsection{Attention-Guided Global Context Module}
The clinical practice and data analysis have demonstrated that the pneumonia lesion areas, including the COVID-19, have no specific distribution characteristics and are generally scattered. Moreover, lung opacities in the CT image are inhomogenous, and there is no clear center or boundary. These undoubtedly pose great challenges to the segmentation model, including: (1) The scattered lesion areas can easily lead to missed detection or incomplete detection; (2) The boundary of the lesion area is not easy to segment accurately and sharply. To address these issues, we introduce more effective and valuable short-connection encoder as guidance. The evolution of encoder features is specifically manifested in two aspects: First, we use spatial attention and boundary attention to filter the features in terms of highlighting the important spatial locations and boundary details. Second, we capture the context dependence of different regions which constrains the generation of more complete and accurate segmentation result. The overall architecture is illustrated in Fig. \ref{fig3}. The AGGC module can not only ensure the integrity of the features of the local lesion area, but also perceive the global correlation between different areas. The specific process is as follows.

Although the encoder features can provide rich details and other useful information for the decoder, it also has much redundancy.
Therefore, the spatial attention \cite{p2-woo2018cbam} is introduced to modify the encoder features from the spatial dimension, focusing on highlighting the important spatial positions. Specifically, we first apply the average-pooling and max-pooling operations along the channel axis on the input features, and then concatenate them to generate an efficient feature descriptor. Finally, a convolution layer with the kernel size of $3\times 3$ followed by a sigmoid function is used to generate a spatial attention map $A^i_s$:
\begin{equation}
A^i_s=\sigma(conv_{3\times3}(concat(avepool(f^i),maxpool(f^i))))
\end{equation}

% \rc{
% \begin{equation}
%     \begin{aligned}
%     &A^i_s=SA(f^i) \\
%     &= \sigma(conv_{3\times3}(concat(avepool(f^i),maxpool(f^i))))
%     \end{aligned}
% \end{equation}
% }
where $avepool$ and $maxpool$ denote the average-pooling and max-pooling along the channel axis, respectively. With the spatial attention map, the initial encoder features $f^i$ can be refined as spatial-enhanced features $f^i_s$ via residual connection:
\begin{equation}
f^i_s=A^i_s\odot f^i+f^i
\end{equation}
where $\odot$ represents element-wise multiplication.
The visualization of spatial-enhanced features is shown in Fig. \ref{fig5}(c). It can be seen that the spatial-enhanced features have a high response in the infected area to be segmented, and a low response in the irrelevant background area, thus achieving the effect of highlighting important spatial locations and suppressing interferences.

As mentioned above, segmenting the lesion areas with clear and definite boundary is very important for diagnosis. In order to maintain the sharpness and clarity of the boundary in each decoding stage, we directly use the boundary map $S_b$ generated by the BA unit to emphasize the important boundary details. To highlight the boundary locations without losing other important information, we still use the residual connection for integration:

\begin{figure}[!t]
    \centerline{\includegraphics[width=\columnwidth]{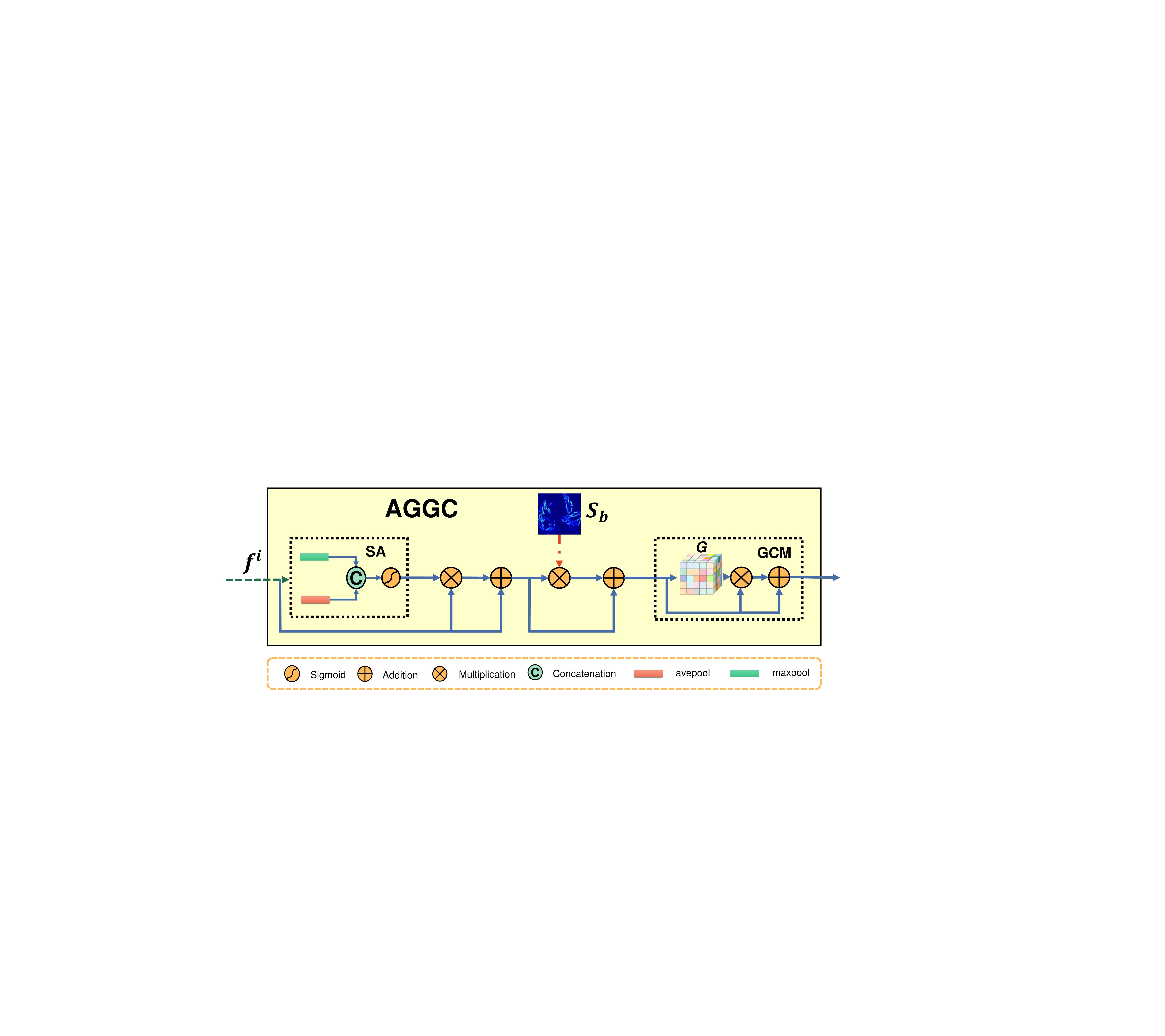}}
    \caption{Illustration of the proposed AGGC module, integrating the important locations, boundary details, and global context. ``SA'' is the spatial attention unit, ``$S_b$'' denotes boundary attention map generated by the BA unit, ``GCM'' is the global context modeling unit that models the global context dependence, and ``G'' is the global context-aware feature map.}
\label{fig3}
\end{figure}

\begin{equation}\label{Eq5}
f^i_b=S_b\odot f^i_s+f^i_s
\end{equation}
where $f^i_b$ represents the boundary-enhanced features, and $S_b$ is the boundary map generated by the BA unit.

In addition to highlighting the important location and boundary of the initial encoder features, we also need to solve the problem of incomplete detection caused by scattered lesions. Therefore, modeling the dependencies between different pixels is an effective solution. In this way, if an infected area is segmented, other scattered interference areas highly related to it may also be located. To this end, we introduce the global context modeling (GCM) unit to achieve feature alignment and mutual enhancement between features by modeling the dependencies between different locations in the feature map. Following \cite{zhang2020dense}, we first learn the global context-aware feature map $G^i$ by measuring the influence of the embedding features of all other spatial locations on the current location:
%In addition to highlighting the important location and boundary of the initial encoder features, we also need to solve the problem of incomplete detection caused by scattered lesions. In other words, the lesion areas have similar appearance characteristics, so how to better correlate them will conduce to obtain complete detection results and reduce the missed detection rate. To this end, we introduce the Global Context Modeling (GCM) unit to achieve feature alignment and mutual reinforcement between features by modeling the dependence relationships among different locations in the feature map. Following \cite{zhang2020dense}, we learn the global context-aware feature map $G^i$ by measuring the influence of the embedding features of all other spatial locations on the current location.
\begin{equation}
    \begin{aligned}
    &G^i = \bigtriangledown (\bigtriangleup (\widetilde{f^i_{b}} \otimes \omega^i  ))\\
    &= \bigtriangledown (\bigtriangleup (\widetilde{f^i_{b}})\otimes Norm(((\bigtriangleup (\widetilde{f^i_{b}}))^T \otimes \bigtriangleup (\widetilde{f^i_{b}}))^T))
    \end{aligned}
\end{equation}
where $\widetilde{f^i_{b}}$~are the normalized boundary-enhanced features, $\bigtriangleup(\cdot)$ reshapes a matrix of $\mathbb{R}^{D_{1}\times D_{2}\times D_{3}}$ into $\mathbb{R}^{D_{1}\times D_{23}}$, $\bigtriangledown(\cdot)$ is the inverse operator of $\bigtriangleup(\cdot)$, $\otimes$ is the matrix multiplication, $\omega^i$ is the global context relationship map, and $Norm$ is the normalization operation. Then, the $G^i$ that encodes the global context relationship is used to further refine the boundary-enhanced features in a residual connection manner:
\begin{equation}
f^i_{aggc}=\varsigma \cdot (G^i\odot f^i_b)+f^i_b
\end{equation}
where $\varsigma$ is a learnable weight parameter that controls the contribution of the global contextual information.

\begin{figure}[!t]
\centerline{\includegraphics[width=\columnwidth]{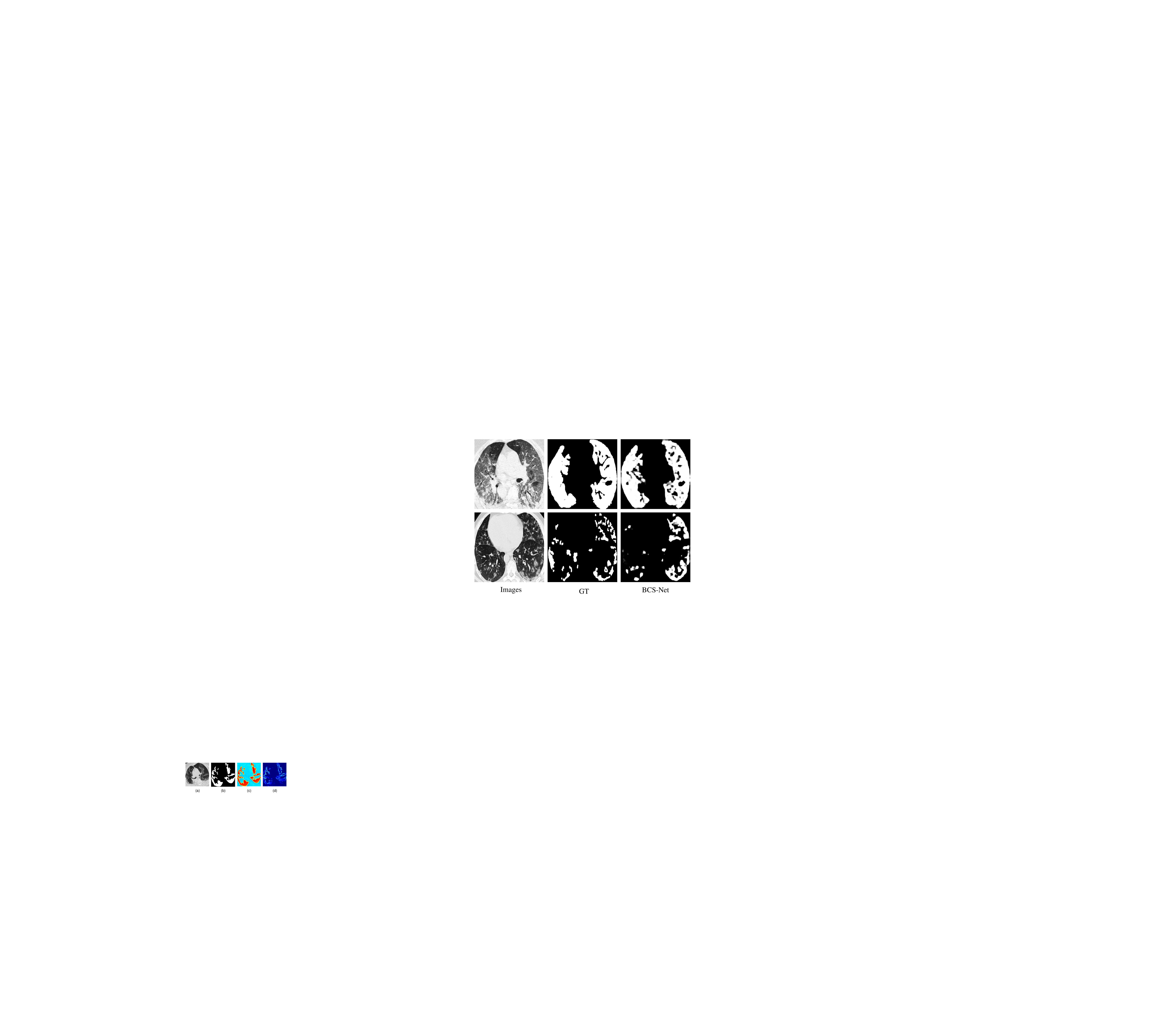}}
\caption{(a) CT Image. (b) Infection segmentation ground truth. (c) Visualization of spatial-enhanced features. (d) Visualization of boundary attention map.}
\label{fig5}
\end{figure}

The entire AGGC model adopts a progressive enhancement structure, which not only highlights important spatial locations and boundary details, but also encodes the context dependencies between different spatial positions. The operation sequence of these three enhancement strategies is spatial attention, boundary attention, and global context. The reasons for this design are as follows: For encoder features with rich information, reducing redundancy is a primary task, otherwise the continued existence of redundancy in the later learning process will weaken the expression of features; Then, in order to ensure that the boundary details are well preserved and highlighted, we apply the boundary attention to spatial-enhanced features; Finally, calculating the global context dependence on relatively clear and distinct features after spatial and boundary enhancements can further strengthen the discrimination of features. This process can obtain encoder features with prominent spatial positions, highlighted boundary details, and comprehensive global context, which can then be used as the guidance for decoder features.

\subsection{Boundary Attention Unit}
For the medical image segmentation task, especially the COVID-19 segmentation where the lesion areas are scattered throughout the image, it is particularly important to clearly and accurately outline the boundaries of the lesion areas for later diagnosis.
If only the semantic information of each pixel is considered and the boundary information is ignored in the segmentation process, this may cause the boundary of the segmentation result to be blurred and not sharp enough.

\begin{figure}[!t]
    \centerline{\includegraphics[width=\columnwidth]{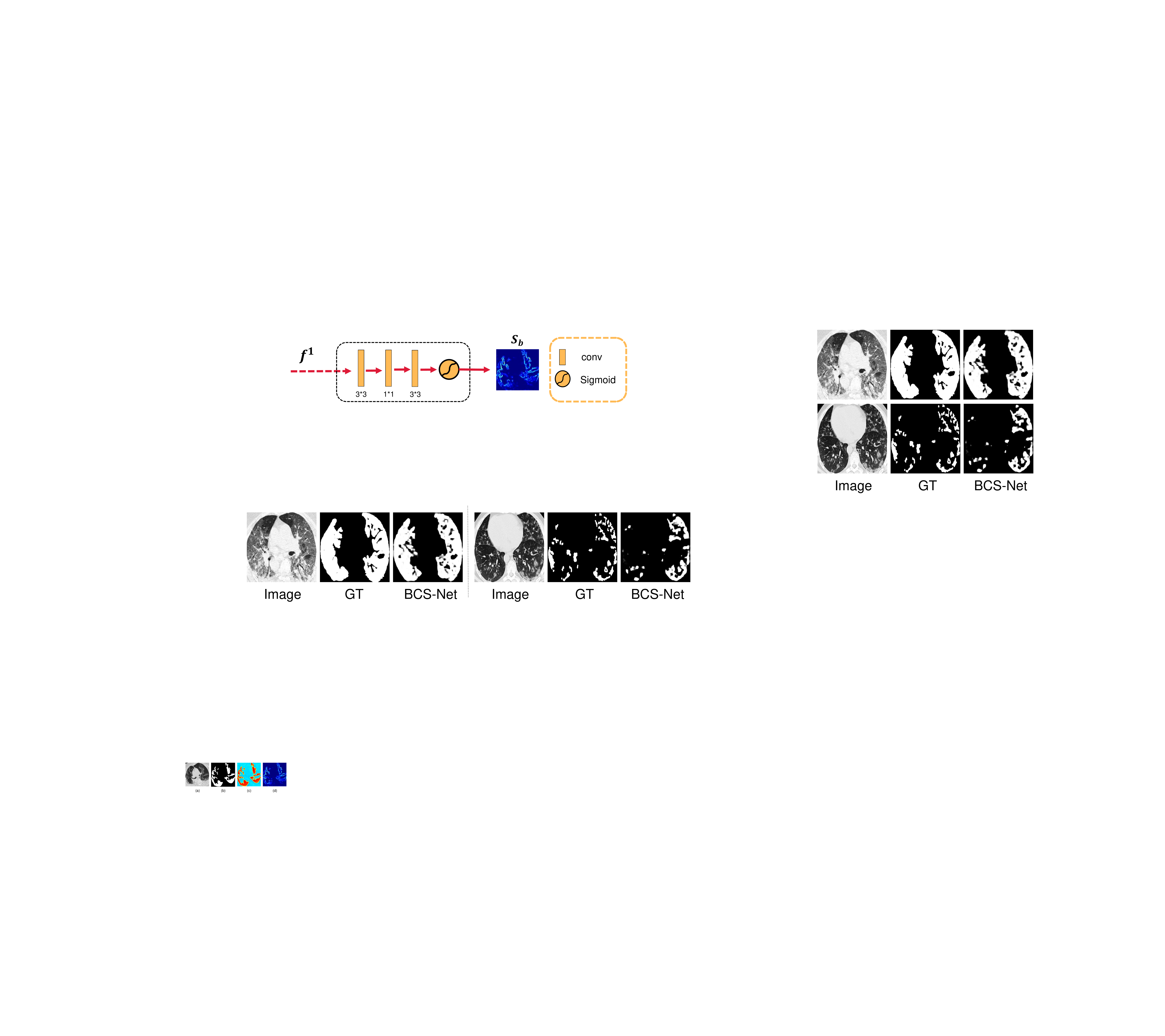}}
    \caption{Illustration of the proposed Boundary Attention (BA) unit, where $S_b$ denotes boundary map.}
\label{fig4}
\end{figure}

%To this end, the validity of the boundary constraint for medical image segmentation has been demonstrated in \cite{zhang2019Etnet,infnet}.
Inspired by \cite{zhang2019Etnet,infnet}, we introduce the boundary constraint into our approach by utilizing the boundary attention unit, as shown in Fig. \ref{fig4}. On the one hand, we also introduce the boundary supervised learning to the network. In this way, the network can gradually learn the features of generating a clear boundary map.
%We use the low-level features $f^1$ from the first encoder layer to generate boundary map.
Specifically, the low-level features $f^1$ are fed into three convolution layers with different kernel sizes, and then a sigmoid function is employed to generate the final boundary attention map:
\begin{equation}
S_b=\sigma(conv_{1\times1}(conv_{3\times3}(conv_{1\times1}(f^1))))
\end{equation}
where $f^1$ are the encoder features of the first layer, $conv_{n \times n}$ denotes the convolution layer with the kernel size of $n\times n$.
The learning process of boundary attention map is a supervised learning manner, which can guarantee the effectiveness and pertinence of learning. As shown in Fig. \ref{fig5}(d), the boundary attention map $S_b$ effectively highlights the boundary positions of the infected regions to be segmented, which can be further used to refine the high-level features in the AGGC module.

In addition to generating boundary map through label supervised learning, we also treat the generated boundary map as an attention weight to highlight the encoder features as introduced in Eq. (\ref{Eq5}). Different from the work \cite{infnet}, we did not use the boundary features to perform feature enhancement. Instead, we adopt a method similar to spatial attention, treat the generated boundary map as an attention map, and use the residual connection for feature enhancement. In this way, boundary constraints can be introduced more intuitively and efficiently, and at the same time, unnecessary information redundancy in the feature space can be reduced. This boundary attention map will be used to enhance all the high-level features (\ie, $f^2$, $f^3$, $f^4$).

\subsection{Semantic Guidance Unit}
The features learned by different encoder layers include different information. For example, the low-level features highlight the details of the scene such as the texture and boundary, while the high-level features tend to learn the semantic and category attributes of the scene. For the segmentation task of COVID-19 lesion areas, there are relatively more background interferences, and how to effectively suppress them is particularly important. Considering the importance of high-level semantic information for suppressing irrelevant background interference, we design a Semantic Guidance (SG) unit to aggregate multi-scale high-level features and formulate a semantic guidance map to refine the decoder features.

For the feature fusion of different scales, the correlation between feature maps of adjacent scales on the feature pyramid is the strongest. However, in the previous multi-scale feature fusion methods, some features are usually up-sampled or down-sampled multiple times, which will lose the semantic information of the original features and even cause semantic ambiguity. Based on this, we try our best to avoid the worthless and unadvisable multiple cross-scale sampling of single layer. As illustrated in Fig. \ref{fig2}, the spatial resolution of high-level features ${f^2,f^3,f^4}$ are unified on the intermediate resolution of the third encoder layer, which not only avoids the information loss due to multiple sampling, but also preserves the correlation between adjacent-scale features to the utmost extent. To be more specific, the encoder features $f^2$ and $f^4$ are $\times2$ down-sampled and $\times2$ up-sampled to the resolution of features $f^3$  respectively, and then they are fused together to obtain a semantic guidance map through the addition and convolution operations. The process can be formulated as:
\begin{equation}
S_s=conv(down_2(f^2)+f^3+up_2(f^4))
\end{equation}
where $up_2$ and $down_2$ are the $\times 2$ spatial up-sampling and down-sampling, respectively.
The learned semantic map $S_s$ encodes the multi-level semantic information, which is served as a global guidance in BCSR block. Since we choose the middle layer as the scale standard of feature output, the semantic map $S_s$ only needs to be down-sampled once in the first BCSR block and up-sampled once in the last BCSR block, respectively. It effectively avoids the multiple cross-level sampling of $S_s$ and retains the feature information of $S_s$ to the maximum extent.

\subsection{Loss Function}
%For the supervision of network, we design a hierarchical loss function by using explicit deep supervision on the side outputs of different scales, which imposes stronger constraints to learn more powerful lung infection region features. In \cite{l1-qin2019basnet, l2-wei2020f3net}, a weighted Intersection over Union (IOU) loss and a weighted binary cross entropy (BCE) loss are used. In addition, inspired by \cite{infnet}, we introduce boundary supervision to measure the dissimilarity between the boundary attention map and the boundary ground truth map via the standard BCE loss.

For the supervision of network, the overall loss function involves two parts: (1) The segmentation loss on three decoder outputs (\ie, $S_2$, $S_3$, and $S_4$) and the semantic loss on the semantic map $S_s$ via weighted IOU loss and weighted BCE loss. Inspired by \cite{l1-qin2019basnet, l2-wei2020f3net}, the weighted IOU loss provides effective global (image level) supervision, and the weighted BCE loss provides effective local (pixel level) supervision. Combining them, accurate segmentation result can be obtained from the image level and pixel level, which is defined as:
\begin{equation}
    l_{s} = \sum_{k=\{S_2,S_3,S_4,S_s\}}(l^k_{wiou}+l^k_{wbce})
\end{equation}
\begin{equation}
    \begin{aligned}
    & l^k_{wbce} = \frac{\sum_{x = 1}^{w}\sum_{y = 1}^{h}-G_{xy}\log(P^k_{xy})\ast (1+\varepsilon_{xy})}{\sum_{x = 1}^{w}\sum_{y = 1}^{h}\varepsilon_{xy}}\\
    & - \frac{\sum_{x = 1}^{w}\sum_{y = 1}^{h}(1-G_{xy})\log(1-P^k_{xy})\ast (1+\varepsilon_{xy})}{\sum_{x = 1}^{w}\sum_{y = 1}^{h}\varepsilon_{xy}}
    \end{aligned}
\end{equation}
\begin{equation}
    l^k_{wiou} =1 - \frac{\sum_{x = 1}^{w}\sum_{y = 1}^{h}(P^k \cap G)\ast (1+\varepsilon_{xy})}{\sum_{x=1}^{w}\sum_{y = 1}^{h}(P^k \cup  G)\ast (1+\varepsilon_{xy})}
\end{equation}
where $l^k_{wiou}$ and $l^k_{wbce}$ denote the weighted IOU loss and weighted BCE loss on different supervision types respectively, $P^k$ is the prediction map (\ie, $S_2$, $S_3$, $S_4$, $S_s$), $G$ is the segmentation ground truth, $w$ and $h$ are the width and height of the input image, $(x, y)$ denotes the coordinate of each pixel in image, and $\varepsilon$ is a weight hyperparameter.
(2) The boundary loss $l^b_{bce}$ on the boundary attention map $S_b$ via standard BCE loss, which is defined as:
\begin{equation}
l^b_{bce} = \sum_{x=1}^{w}\sum_{y=1}^{h}-G^b_{xy}\log(P^b_{xy})-(1-G^b_{xy})\log(1-P^b_{xy})
\end{equation}
where $G^b$ is the boundary ground truth, and $P^b$ is the predicted boundary map.

Therefore, the overall loss function can be defined as:
\begin{equation}
\iota_{total} = l_{s}+l^b_{bce}
\end{equation}

\section{EXPERIMENTS}
\label{sec:guidelines}

\subsection{Datasets and Evaluation Metrics}
At present, there are few public COVID-19 lung CT datasets for infection segmentation.
In order to have relatively sufficient samples for training, we merged the two public datasets \cite{segdata1,segdata2} to obtain $1018$ high-quality CT images, and further divided them into $719$ training images and $300$ testing  images.
For each CT slice, it contains the corresponding infection mask, and the infection edge obtained from infection mask using the Canny operator.
Note that, the COVID-19 classification aims to determine whether lung CT images are infected with COVID-19, and the datasets used should contain normal and abnormal data. In contrast, the purpose of COVID-19 infection segmentation is to locate the infected area accurately and completely, so the default option is that the infected area is present in every CT image of the used dataset. In other words, consistent with existing methods \cite{2022SSA,infnet,COPLENet}, all images in the dataset \cite{segdata1,segdata2} we use are COVID-19 infection data.

\begin{figure}[!t]
    \centerline{\includegraphics[width=\columnwidth]{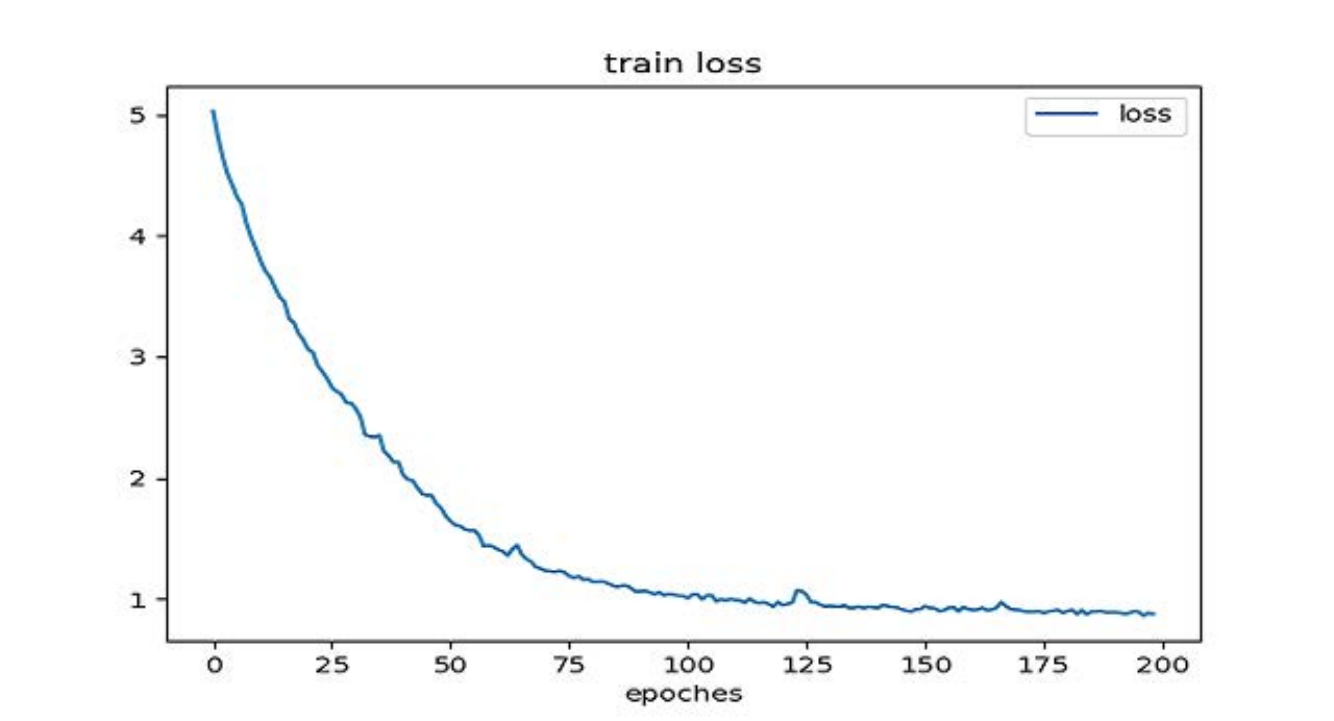}}
    \caption{The training loss curve.}
    \label{fig7}
\end{figure}

Following \cite{infnet}, six metrics are employed for quantitative evaluation, including Precision ($Prec$), Recall ($Recall$), Dice Similarity Coefficient ($DSC$) \cite{2020Lung}, Structure Measure ($S_m$) \cite{S}, Enhance-alignment Measure ($E_\phi$) \cite{E}, and Mean Absolute Error ($MAE$) \cite{crm2019tcsvt}.
%Dice Similarity Coefficient ($DSC$) is a spatial overlap index and a reproducibility validation metric, which evaluates the overlap ratio between prediction map and ground truth map.
%Precision ($Prec$) measures the accuracy of model in classifying a sample as positive, and Recall ($Recall$) measures the proportion of positives that are correctly identified as positive.
The Precision ($Prec$) calculates the ratio of true positive cases in the results predicted as positive cases, and the Recall ($Recall$) score only focuses on the rate that the real label is correctly predicted, which can be computed as:
%Structure Measure ($S_m$) calculates the region-aware structural similarity and object-aware structural similarity between prediction map and ground-truth mask.
%Enhance-alignment Measure ($E_\phi$) captures the global statistics and local pixel matching information to account for both pixel-level and image-level properties.
%Mean Absolute Error ($MAE$) measures the pixel-wise error between prediction map and ground truth.
\begin{equation}
Prec = \frac{TP}{TP + FP}, ~~~Recall = \frac{TP}{TP + FN}
\end{equation}
where $TP$, $FP$, and $FN$ are the true positive, false positive, and false negative, respectively.

Dice Similarity Coefficient ($DSC$) is a spatial overlap index and a reproducibility validation metric, which evaluates the overlap ratio between prediction map ($P$) and ground truth map ($G$):
\begin{equation}
DSC = \frac{2\left\lvert P\bigcap G\right\rvert }{\left\lvert P \right\rvert + \left\lvert G \right\rvert }
\end{equation}

Structure Measure ($S_m$) calculates the similarity between prediction map and ground-truth mask, including a region-aware structural similarity measure $S_o$ and an object-aware structural similarity measure $S_r$, which is
calculated as:
\begin{equation}
S_m = \alpha \cdot S_o(P, G) + (1 - \alpha) \cdot S_r(P, G)
\end{equation}
where $\alpha \in [0,1]$, we set $\alpha = 0.5$ in our implementation suggested in the original paper.

Enhance-alignment Measure ($E_\phi$) captures global statistics and local pixel matching information to account for both pixel-level and image-level properties, which is defined as:
\begin{equation}
E_\phi = \frac{1}{w \times h}\sum_{x = 1}^{w}\sum_{y = 1}^{h}\phi (x, y)
\end{equation}
where $h$ and $w$ are the height and the width of the input image, and $(x,y)$ denotes the coordinate of each pixel in prediction map and ground truth.

\begin{figure*}[!t]
\centerline{\includegraphics[width=1\textwidth]{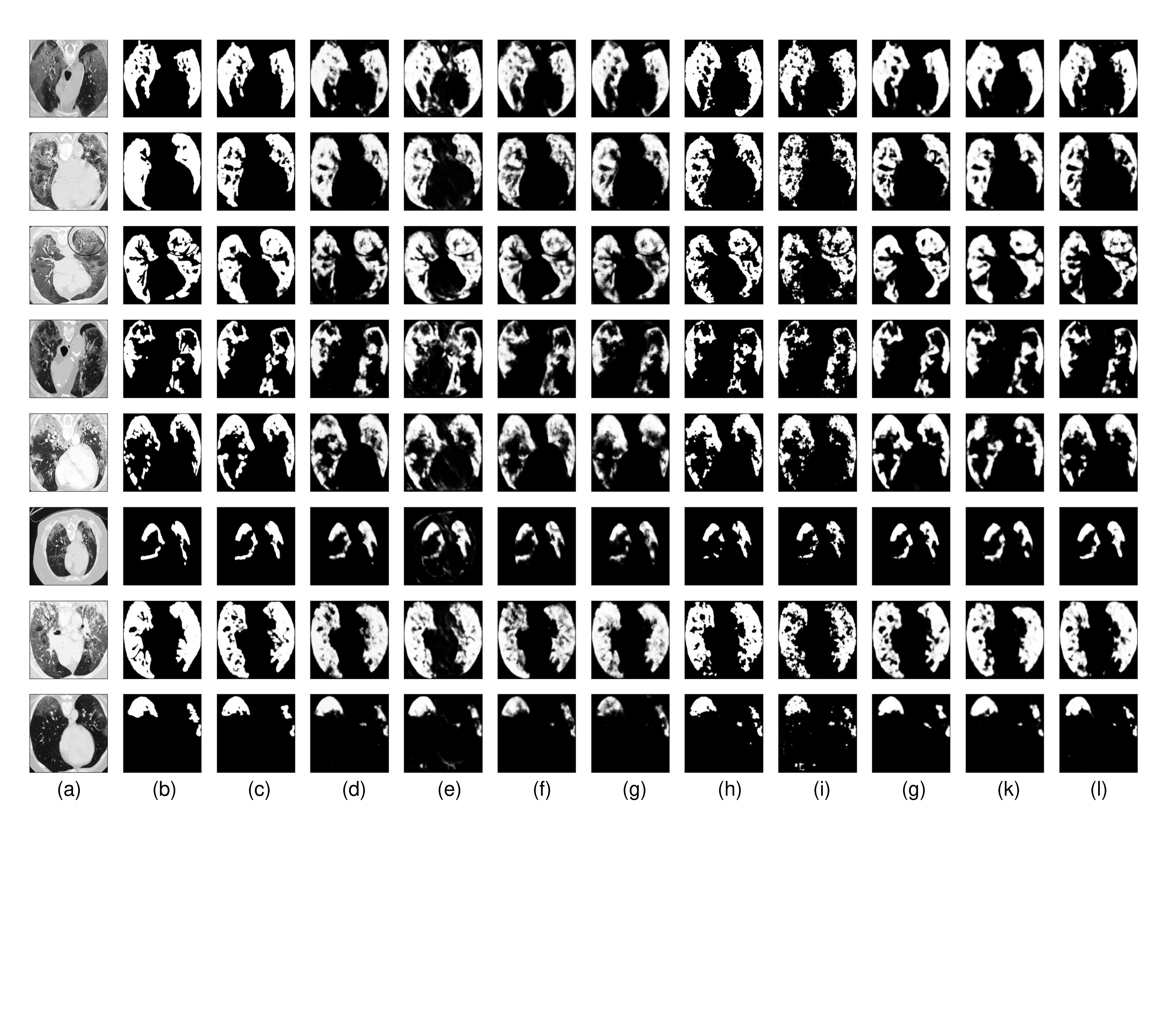}}
\caption{Some visual examples of different methods. (a) Input image. (b) Ground truth. (c)-(l) Results of our method, UNet \cite{unet}, Resnet34-UNet \cite{unet}, UNet++ \cite{unet++}, Attention-UNet \cite{attentionunet}, CE-Net \cite{Cenet},  FCN \cite{FCN}, Inf-Net \cite{infnet}, AnamNet \cite{AnamNet}, and CopleNet \cite{COPLENet}.
%We can see that our method (BCS-Net) exhibits competitive visual effects in both the segmentation ability of the infected regions and the suppression ability of the background interferences.
}
\label{fig6}
\end{figure*}

Mean Absolute Error ($MAE$) measures the errors between prediction map and ground truth:
\begin{equation}
MAE = \frac{1}{w \times h}\sum_{x = 1}^{w}\sum_{y = 1}^{h}\left\lvert P(x, y) - G(x, y)\right\rvert
\end{equation}

For these six measurements, in addition to the MAE score, all other indicators are that the larger the value, the better the performance.

\subsection{Training Strategies and Implementation Details}
We implement the proposed network via the PyTorch toolbox and is accelerated by an NVIDIA GeForce RTX $2080$Ti GPU. We also implement our network by using the MindSpore Lite tool\footnote{\url{https://www.mindspore.cn/}}.
The Res2Net-50 \cite{Res2Net} pretrained on ImageNet \cite{imagenet} is employed as the backbone feature extractor in the experiment.
We have two points in particular to note: (1) The fundamental reason we choose Res2Net-50 as the backbone network is for fair comparison, so we follow the setting of Inf-Net \cite{infnet}. In addition, its technical advantages (\eg, hierarchical residual-like connections within one single residual block, multi-scale features at a granular level, and increased range of receptive fields) and low computational cost are also the reasons for our choice. (2) The way we load the pretrained ImageNet data is consistent with existing work \cite{infnet,DBLP:conf/wacv/LaradjiRMLLKP0N21}. Specifically, it is well known that CT images are grayscale, and when we load the CT image, we still use the RGB three-channel method, but the corresponding pixel values in the three channels are the same.
In order to computational efficiency, we resize all the training and testing images to $352\times352$.
The batch size is $8$ and the training process converges until $200$ iterations. We use Adam \cite{adam} as a learning rate optimizer with betas of $(0.9, 0.999)$, and the learning rate is set to $3e^{-4}$.
The training loss curve of our network is shown in Fig. \ref{fig7}.
The average inference time of our method is $0.036$ second ($27$ FPS) for processing an image with the size of $352\times352$, the GFLOPs is $28.02$, and the parameters are $44,822,580$. The code and results of our BCS-Net can be found from the link of \url{https://github.com/rmcong/BCS-Net-TIM22}.

\subsection{Comparison with State-of-the-art Method}\label{formats}
We compare our method with some state-of-the-art methods, including five classical segmentation models (UNet \cite{unet}, UNet++ \cite{unet++}, Attention-UNet \cite{attentionunet}, Resnet34-UNet \cite{unet}, CE-Net \cite{Cenet}, and FCN \cite{FCN}), and three state-of-the-art segmentation methods for COVID-19 (Inf-Net \cite{infnet}, CopleNet \cite{COPLENet}, and AnamNet \cite{AnamNet}).
In order to ensure the fairness of the experiment, all the comparison methods are retrained on the same dataset under the default parameter settings.

\subsubsection{Qualitative Comparison}
Fig. \ref{fig6} shows the qualitative comparisons of our model with the other state-of-the-art methods.
It can be seen that the results of our method have greater advantages in terms of detection accuracy, completeness, and sharpness.
For example, in the first image, the general medical image segmentation networks (\eg, UNet \cite{unet}, UNet++ \cite{unet++}, Attention-UNet \cite{attentionunet}, Resnet34-UNet \cite{unet}) often cannot effectively suppress the interference of background regions, such as the areas between the two lungs. In contrast, the detection results of the COVID-19 segmentation network are better, but the existing methods (\eg, Inf-Net \cite{infnet}, CopleNet \cite{COPLENet}, and AnamNet \cite{AnamNet}) cannot completely suppress these interferences. For example, the area under the left lung is not effectively suppressed. Our proposed method has better performance in these aspects, and has a stronger ability to detect details. The last image also confirms that our proposed method can accurately detect infected areas and suppress irrelevant background areas. In addition, our method has more complete structure and sharper boundaries. In the third image, the existing methods (\eg, Inf-Net \cite{infnet}, CopleNet \cite{COPLENet}, and AnamNet \cite{AnamNet}) cannot completely and continuously detect the infected regions at the bottom of the left lung, while our method can detect them clearly, accurately, and completely. Moreover, compared with the existing methods (\eg, UNet++ \cite{unet++}, CopleNet \cite{COPLENet}, Attention-UNet \cite{attentionunet}), our results have clearer boundaries, which are very important for doctors to diagnose and treat. In general, our method exhibits competitive visual effects in both the segmentation ability of the infected regions and the suppression ability of the background regions.

\begin{table}
    \centering
    \caption{Quantitative Comparisons with Different Methods. The best performance is marked in bold.}
     \renewcommand{\arraystretch}{1.2}
     \setlength{\tabcolsep}{1mm}{
    \begin{tabular}{|c|c|c|c|c|c|c|}
    \hline
    Methods&$DSC \uparrow$&$Prec \uparrow$&$Recall \uparrow$&$S_m \uparrow$&$E_\phi \uparrow$&$MAE \downarrow$\\
    \hline
    UNet&0.777&0.804&0.814&0.862&0.917&0.020\\
    \hline
    UNet++&0.771&0.836&0.780&0.867&0.906&0.021\\
    \hline
    Attention\_UNet&0.746&0.818&0.768&0.853&0.913&0.021\\
    \hline
    Resnet34\_UNet&0.720&0.702&0.836&0.812&0.873&0.030\\
    \hline
    FCN&0.800&0.839&0.791&0.855&0.949&0.020\\
    \hline
    CE-Net&0.818&0.834&0.824&0.854&0.960&0.017\\
    \hline
    AnamNet&0.775&0.831&0.776&0.856&0.920&0.021\\
    \hline
    CopleNet&0.816&\textbf{0.850}&0.821&0.874&0.944&0.016\\
    \hline
    Inf-Net&0.828&0.831&0.846&0.877&0.963&0.016\\
    \hline
    BCS-Net [ours]&\textbf{0.844}&0.841&\textbf{0.861}&\textbf{0.880}&\textbf{0.972}&\textbf{0.015}\\
    \hline
    \end{tabular}
     }
    \label{table:1}
\end{table}

\subsubsection{Quantitative Comparison}
The quantitative comparisons are reported in Table \ref{table:1}. In addition to the Precision score, our proposed BCS-Net achieves the best performance in other five measurements on the testing dataset.
Compared with the classical segmentation models (\eg, UNet \cite{unet}, UNet++ \cite{unet++}, Attention-UNet \cite{attentionunet}, Resnet34-UNet \cite{unet}), the well-designed methods for COVID-19 are more competitive. For example, compared with the UNet \cite{unet}, the percentage gain of the Inf-Net \cite{infnet} reaches 6.6\% for $DSC$ score, and the percentage gain of our method is 8.6\%. For the $E_\phi$, the performance improvement is more obvious. Specifically, the performance of Inf-Net \cite{infnet} is increased by 5.7\% compared with the UNet++ \cite{unet++}, and the performance is increased by 9.0\% compared with the Resnet34-UNet \cite{unet}.
For the infection segmentation method for COVID-19, our proposed BCS-Net achieves better performance. For example, compared with the \emph{second best} method, the percentage gain of $DSC$ is 1.9\%, the $Recall$ score is 1.8\%, and the $E_\phi$ score achieves 1.0\%. In general, our detection effect is superior in quantitative measurements.

\subsection{Ablation Study}
\begin{figure}[!t]
    \centering
    \centerline{\includegraphics[width=\columnwidth]{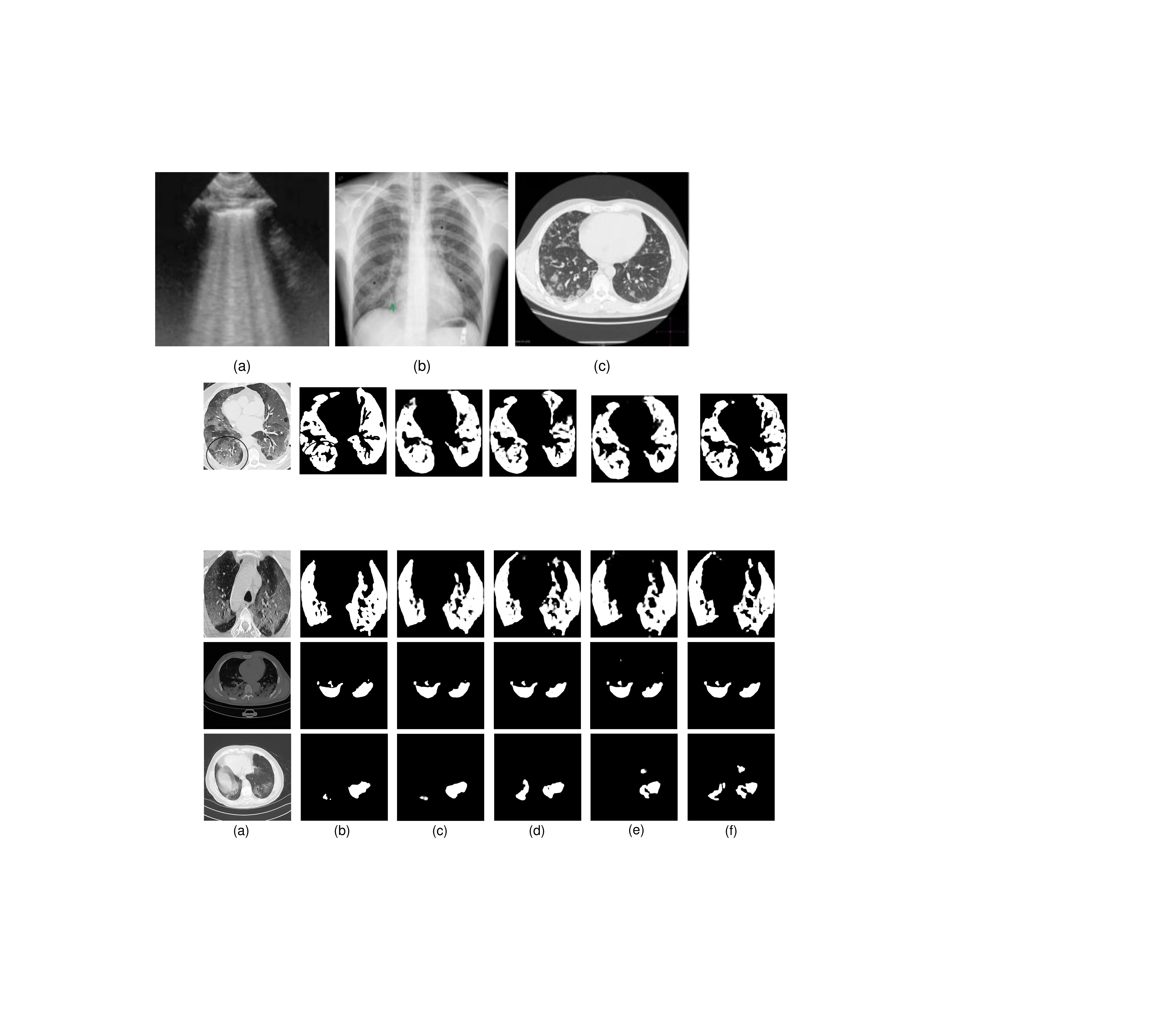}}
    \caption{Visual comparison of BCS-Net variants equipped with different modules. (a) Images. (b) Ground truth. (c) BCS-Net. (d) w/o AGGC. (d) w/o SG. (f) w/o SG w/ PPD.}
    \label{fig8}
\end{figure}

We conduct several ablation experiments to validate the performance and effectiveness of two contributed components of our proposed model, including Attention-Guided Global Context (AGGC) module and Semantic Guidance (SG) unit. The results are given in Table \ref{table:2} and Fig. \ref{fig8}.

\begin{table}
    \centering
    \caption{Quantitative evaluation results of different modules used in our proposed model.}
         \renewcommand{\arraystretch}{1.2}
     \setlength{\tabcolsep}{1mm}{
    \begin{tabular}{|c|c|c|c|c|c|c|c|}
    \hline
    No. &Variations&$DSC \uparrow$&$Prec \uparrow$&$Recall \uparrow$&$S_m \uparrow$&$E_\phi \uparrow$&$MAE \downarrow$\\
    \hline
    1& BCS-Net&0.844&0.841&0.861&0.880&0.972&0.015\\
    \hline
    2& w/o AGGC&0.838&0.835&0.856&0.874&0.970&0.015\\
    \hline
    3& w/o SG&0.837&0.840&0.850&0.872&0.969&0.016\\
    \hline
    4& \makecell[*l]{w/o SG \\ w/ PPD} &0.836&0.838&0.849&0.872&0.969&0.016\\
    \hline
    \end{tabular}
     }
    \label{table:2}
\end{table}

Firstly, we verify the effect of AGGC module on the whole network. The main function of the AGGC module is to strengthen the important spatial position information and correlate the dependencies of different pixels, thereby ensuring a more accurate and complete detection.
We remove the AGGC module, including the guidance of spatial attention and boundary attention, and the global context awareness. In other words, in the BCSR block, the decoding features of the previous layer will not enter the AGGC module but will be passed directly backwards.
Compared with the No.1 and No.2 in Table \ref{table:2}, without the AGGC module, the five indicators fell, and the MAE score remained flat.
As shown in Fig. \ref{fig8}(d), after removing the AGGC module, there are unsuppressed interference noises in the upper left corner, upper right corner and lower right corner of the first image, and the boundaries are blurred. In the second image, the small lesions on the upper right after removing the AGGC module are not detected, while in the third image, there are obvious false detections in the left region. These all demonstrate the effectiveness of our AGGC module.

Secondly, in order to discuss the role of the SG unit, we design two ablation experiments. One is that we directly delete the SG unit, that is, without introducing global semantic information guidance. The other is that we replace the proposed SG unit with the PPD module in Inf-Net \cite{infnet}. The related results are reported in No.3 and No.4 of Table \ref{table:2}.
Compared with the full model, all indicators will be reduced after removing or replacing the SG unit.
In addition, we also found that the complex PPD model has three indicators (\ie, $DSC$, $Prec$, and $Recall$) that it is not as good as removing the SG unit directly. Comparing the SG and PPD modules of introducing global semantic guidance (\ie, No.1 and No.4), we can see that the designed SG model achieves better performance.
After removing the SG module, the upper left and upper right regions in the first image of Fig. \ref{fig8}(e), the upper left region of the second image, and the upper right region of the third image have some irrelevant interferences that are not effectively suppressed. Replacing the SG module with the PPD module also does not improve the performance, and both the left region and the upper right background region in the third image are incorrectly detected as infected regions. Both ablation experiments illustrate the role of the SG module designed in this paper.

\begin{table}
    \centering
    \caption{The performance comparisons of outputs $S_1$, $S_2$ and $S_3$.}
         \renewcommand{\arraystretch}{1.2}
     \setlength{\tabcolsep}{1mm}{
    \begin{tabular}{|c|c|c|c|c|c|c|c|}
    \hline
    No. &Output&$DSC \uparrow$&$Prec \uparrow$&$Recall \uparrow$&$S_m \uparrow$&$E_\phi \uparrow$&$MAE \downarrow$\\
    \hline
    1& $S_1$&0.839&0.851&0.846&0.867&0.965&0.017\\
    \hline
    2& $S_3$&0.802&0.776&0.855&0.853&0.954&0.017\\
    \hline
    3& $S_2$[ours]&0.844&0.841&0.861&0.880&0.972&0.015\\
    \hline
    \end{tabular}
     }
    \label{table:3}
\end{table}
The final output of the network is derived from features at the second decoding stage mainly based on the following two points. First, we design the AGGC module to make full use of the features of the first encoder layer ($f^1$) to supplement boundary information for high-level encoder features ($f^2$, $f^3$ and $f^4$). In implementation, the features $f^1$ are combined with $f^2$, $f^3$ and $f^4$  through the AGGC module respectively. Under such a model framework, we do not perform the AGGC module in the last decoding layer, so the network does not set the output of $S_1$. Second, as we all know, generating the final map at a lower resolution from a higher decoding stage will consume less computational resources, which is beneficial for computational efficiency. Moreover, it has been experimentally verified that its performance will not be obviously degraded. Table \ref{table:3} presents the final segmentation results obtained by different decoding layers. It can be found that in general, $S_2$ achieves better performance than $S_1$ and $S_3$, which also illustrates the effectiveness of our setup. Based on these observations and consideration, the output from the second decoding layer is used as the final segmentation result.

\subsection{Failure Cases and Future Work}
\begin{figure}[!t]
\centering
\centerline{\includegraphics[width=1\columnwidth]{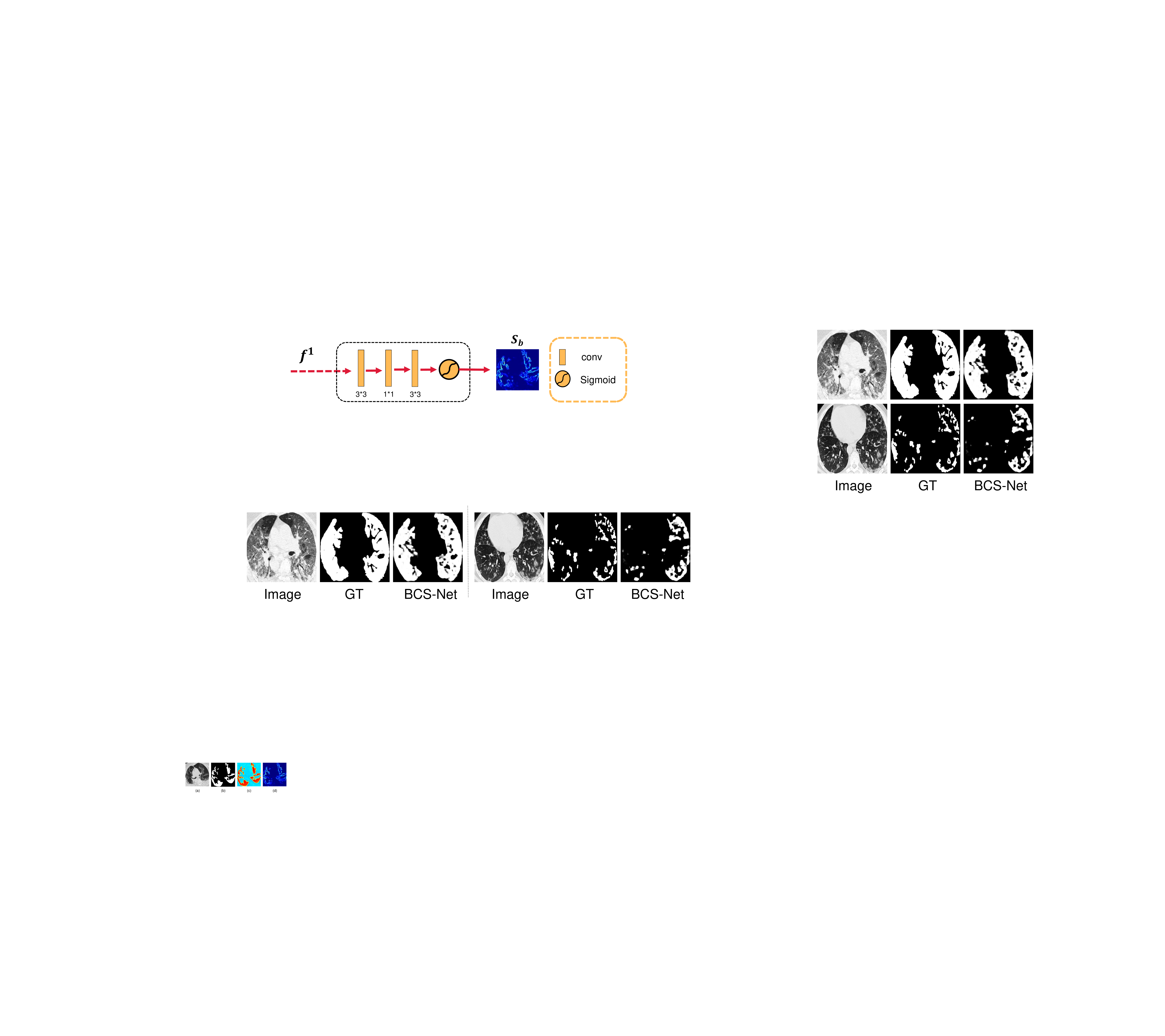}}
\caption{Failure cases of our BCS-Net.}
\label{fig9}
\end{figure}

Compared with other methods, our algorithm achieves a more complete structure and more accurate details, but there is still a certain gap with segmentation ground truth, which we need to solve in the future. Some failure cases are shown in Fig. \ref{fig9}. It can be seen that the segmentation results of our method can maintain the structural integrity in the case of a large area of infection in the whole lung, but there will be some wrongly detected holes, as shown in the left image. In addition, our method also misses some extremely dispersed infected areas. For example, in the right image, many infected areas on the left are not detected, and some scattered infected areas on the right are mistakenly merged together. In the future, we can start from the following aspects to further improve performance. On the one hand, some popular and powerful backbones can be used in place of our basic feature extractor, such as ViT \cite{dosovitskiy2020image} and Swin Transformer \cite{liu2021swin}, which can be used to model long-term dependencies in data through self-attention mechanism. On the other hand, the growth of training data can effectively promote the improvement of model performance and robustness. Therefore, in the future, we call on global scientific research institutions to open source relevant data in accordance with relevant regulations, so as to further promote the development and performance improvement of this field.

\section{Conclusion}
This paper focuses on COVID-19 lung infection segmentation task and proposes an end-to-end framework dubbed as BCS-Net to achieve this.
The proposed network follows an encoder-decoder architecture equipped with Boundary-Context-Semantic Reconstruction (BCSR) blocks, which considers the boundary refinement, context modeling, and semantic constraint jointly.
The attention-guided global context (AGGC) module is designed to select the most valuable encoder features from the perspective of important spatial/boundary locations and context dependence.
Moreover, we design a new semantic guidance (SG) unit to provide the semantic guidance and suppress the background noises by aggregating multi-scale high-level features at the intermediate resolution.
Extensive experiments and ablation studies demonstrate the effectiveness of the proposed BCS-Net architecture.

\par
\ifCLASSOPTIONcaptionsoff
  \newpage
\fi
{
\bibliographystyle{IEEEtran}
\bibliography{ref}

\begin{thebibliography}{10}
\providecommand{\url}[1]{#1}
\csname url@samestyle\endcsname
\providecommand{\newblock}{\relax}
\providecommand{\bibinfo}[2]{#2}
\providecommand{\BIBentrySTDinterwordspacing}{\spaceskip=0pt\relax}
\providecommand{\BIBentryALTinterwordstretchfactor}{4}
\providecommand{\BIBentryALTinterwordspacing}{\spaceskip=\fontdimen2\font plus
\BIBentryALTinterwordstretchfactor\fontdimen3\font minus
  \fontdimen4\font\relax}
\providecommand{\BIBforeignlanguage}[2]{{%
\expandafter\ifx\csname l@#1\endcsname\relax
\typeout{** WARNING: IEEEtran.bst: No hyphenation pattern has been}%
\typeout{** loaded for the language `#1'. Using the pattern for}%
\typeout{** the default language instead.}%
\else
\language=\csname l@#1\endcsname
\fi
#2}}
\providecommand{\BIBdecl}{\relax}
\BIBdecl

\bibitem{boundariesatten}
D.~Mishra, S.~Chaudhury, M.~Sarkar, and A.~S. Soin, ``Ultrasound image
  segmentation: A deeply supervised network with attention to boundaries,''
  \emph{IEEE Trans. Biomed. Eng.}, vol.~66, no.~6, pp. 1637--1648, 2018.

\bibitem{Ultrasound}
G.~Soldati \emph{et~al.}, ``On lung ultrasound patterns specificity in the
  management of {COVID‐19} patients,'' \emph{J. Ultras. Med.}, vol.~39, pp.
  2283--2284, 2020.

\bibitem{r6}
I.~D. Apostolopoulos and T.~A. Mpesiana, ``{COVID-19}: Automatic detection from
  {X-ray} images utilizing transfer learning with convolutional neural
  networks,'' \emph{Phys. Eng. Sci. Med.}, vol.~43, no.~2, pp. 635--640, 2020.

\bibitem{r9}
N.~E.~M. Khalifa, F.~Smarandache, G.~Manogaran, and M.~Loey, ``A study of the
  neutrosophic set significance on deep transfer learning models: An
  experimental case on a limited {COVID-19} chest {X-ray} dataset,'' \emph{Cog.
  Comput.}, pp. 1--10, 2021.

\bibitem{r10}
A.~Abbas, M.~M. Abdelsamea, and M.~M. Gaber, ``Classification of {COVID-19} in
  chest {X-ray} images using {DeTraC} deep convolutional neural network,''
  \emph{Appl. Intell.}, vol.~51, no.~2, pp. 854--864, 2021.

\bibitem{r17}
T.~Zhou, S.~Canu, and S.~Ruan, ``An automatic {COVID-19} {CT} segmentation
  network using spatial and channel attention mechanism,'' \emph{arXiv preprint
  arXiv:2004.06673}, 2020.

\bibitem{wang2020weakly}
X.~Wang \emph{et~al.}, ``A weakly-supervised framework for {COVID-19}
  classification and lesion localization from chest {CT},'' \emph{IEEE Trans.
  Med. Imag.}, vol.~39, no.~8, pp. 2615--2625, 2020.

\bibitem{infnet}
D.-P. Fan \emph{et~al.}, ``{Inf-Net}: Automatic {COVID-19} lung infection
  segmentation from {CT} images,'' \emph{IEEE Trans. Med. Imag.}, vol.~39,
  no.~8, pp. 2626--2637, 2020.

\bibitem{AnamNet}
N.~Paluru \emph{et~al.}, ``{Anam-Net}: Anamorphic depth embedding-based
  lightweight {CNN} for segmentation of anomalies in {COVID-19} chest {CT}
  images,'' \emph{IEEE Trans. Neural Netw. Learn. Syst.}, vol.~32, no.~3, pp.
  932--946, 2021.

\bibitem{COPLENet}
G.~Wang \emph{et~al.}, ``A noise-robust framework for automatic segmentation of
  {COVID-19} pneumonia lesions from {CT} images,'' \emph{IEEE Trans. Med.
  Imag.}, vol.~39, no.~8, pp. 2653--2663, 2020.

\bibitem{r20}
Z.~Han \emph{et~al.}, ``Accurate screening of {COVID-19} using attention-based
  deep {3D} multiple instance learning,'' \emph{IEEE Trans. Med. Imag.},
  vol.~39, no.~8, pp. 2584--2594, 2020.

\bibitem{DBLP:journals/tim/Roy22}
K.~Roy, D.~Banik, D.~Bhattacharjee, O.~Krejcar, and C.~Kollmann, ``{LwMLA-NET}:
  A lightweight multi-level attention-based network for segmentation of
  {COVID-19} lungs abnormalities from {CT} images,'' \emph{IEEE Trans. Instrum.
  Meas.}, vol.~71, pp. 1--13, 2022.

\bibitem{2022SSA}
X.~Wang \emph{et~al.}, ``{SSA-Net}: Spatial self-attention network for
  {COVID-19} pneumonia infection segmentation with semi-supervised few-shot
  learning,'' \emph{Med. Imag. Anal.}, vol.~79, pp. 1--13, 2022.

\bibitem{crm2020tc}
C.~Li, R.~Cong, S.~Kwong, J.~Hou, H.~Fu, G.~Zhu, D.~Zhang, and Q.~Huang,
  ``{ASIF-Net}: Attention steered interweave fusion network for {RGB-D} salient
  object detection,'' \emph{IEEE Trans. Cybern.}, vol.~50, no.~1, pp. 88--100,
  2021.

\bibitem{crmbridgenet}
Q.~Tang, R.~Cong, R.~Sheng, L.~He, D.~Zhang, Y.~Zhao, and S.~Kwong,
  ``Bridgenet: {A} joint learning network of depth map super-resolution and
  monocular depth estimation,'' in \emph{Proc. ACM MM}, 2021, pp. 2148--2157.

\bibitem{crmunderwater}
C.~Li, S.~Anwar, J.~Hou, R.~Cong, C.~Guo, and W.~Ren, ``Underwater image
  enhancement via medium transmission-guided multi-color space embedding,''
  \emph{{IEEE} Trans. Image Process.}, vol.~30, pp. 4985--5000, 2021.

\bibitem{crmDPANet}
Z.~Chen, R.~Cong, Q.~Xu, and Q.~Huang, ``{DPANet}: Depth potentiality-aware
  gated attention network for {RGB-D} salient object detection,'' \emph{IEEE
  Trans. Image Process.}, vol.~30, pp. 7012--7024, 2021.

\bibitem{crmglnet}
R.~Cong, N.~Yang, C.~Li, H.~Fu, Y.~Zhao, Q.~Huang, and S.~Kwong,
  ``Global-and-local collaborative learning for co-salient object detection,''
  \emph{IEEE Trans. Cybern.}, early access, doi: 10.1109/TCYB.2022.3169431.

\bibitem{crmRRNet}
R.~Cong, Y.~Zhang, L.~Fang, J.~Li, Y.~Zhao, and S.~Kwong, ``{RRNet}: Relational
  reasoning network with parallel multi-scale attention for salient object
  detection in optical remote sensing images,'' \emph{IEEE Trans. Geosci.
  Remote Sens.}, vol.~60, pp. 1558--0644, 2022.

\bibitem{crm2022rsi}
X.~Zhou, K.~Shen, L.~Weng, R.~Cong, B.~Zheng, J.~Zhang, and C.~Yan,
  ``Edge-guided recurrent positioning network for salient object detection in
  optical remote sensing images,'' \emph{IEEE Trans. Cybern.}, 2022.

\bibitem{crmDBLP:journals/spl/HuJCGS21}
J.~Hu, Q.~Jiang, R.~Cong, W.~Gao, and F.~Shao, ``Two-branch deep neural network
  for underwater image enhancement in {HSV} color space,'' \emph{{IEEE} Signal
  Process. Lett.}, vol.~28, pp. 2152--2156, 2021.

\bibitem{crmDBLP:journals/tip/LiAHCGR21}
C.~Li, S.~Anwar, J.~Hou, R.~Cong, C.~Guo, and W.~Ren, ``Underwater image
  enhancement via medium transmission-guided multi-color space embedding,''
  \emph{{IEEE} Trans. Image Process.}, vol.~30, pp. 4985--5000, 2021.

\bibitem{crmDBLP:journals/tip/WenYZCSZZBD21}
H.~Wen, C.~Yan, X.~Zhou, R.~Cong, Y.~Sun, B.~Zheng, J.~Zhang, Y.~Bao, and
  G.~Ding, ``Dynamic selective network for {RGB-D} salient object detection,''
  \emph{{IEEE} Trans. Image Process.}, vol.~30, pp. 9179--9192, 2021.

\bibitem{crmDBLP:journals/tmm/MaoJCGSK22}
Y.~Mao, Q.~Jiang, R.~Cong, W.~Gao, F.~Shao, and S.~Kwong, ``Cross-modality
  fusion and progressive integration network for saliency prediction on
  stereoscopic 3d images,'' \emph{{IEEE} Trans. Multim.}, vol.~24, pp.
  2435--2448, 2022.

\bibitem{crmCoADNet}
Q.~Zhang, R.~Cong, J.~Hou, C.~Li, and Y.~Zhao, ``{CoADNet}: Collaborative
  aggregation-and-distribution networks for co-salient object detection,'' in
  \emph{Proc. NeurIPS}, 2020, pp. 6959--6970.

\bibitem{crmACMMM20-1}
P.~Wen, R.~Yang, Q.~Xu, C.~Qian, Q.~Huang, R.~Cong, and J.~Si, ``{DMVOS}:
  Discriminative matching for real-time video object segmentation,'' in
  \emph{Proc. ACM MM}, 2020, pp. 2048--2056.

\bibitem{crm-nc}
C.~Li, R.~Cong, C.~Guo, H.~Li, C.~Zhang, F.~Zheng, and Y.~Zhao, ``A parallel
  down-up fusion network for salient object detection in optical remote sensing
  images,'' \emph{Neurocomputing}, vol. 415, pp. 411--420, 2020.

\bibitem{crm2019tgrs}
C.~Li, R.~Cong, J.~Hou, S.~Zhang, Y.~Qian, and S.~Kwong, ``Nested network with
  two-stream pyramid for salient object detection in optical remote sensing
  images,'' \emph{IEEE Trans. Geosci. Remote Sens.}, vol.~57, no.~11, pp.
  9156--9166, 2019.

\bibitem{crm-acmmm}
D.~Jing, S.~Zhang, R.~Cong, and Y.~Lin, ``Occlusion-aware bi-directional guided
  network for light field salient object detection,'' in \emph{Proc. {ACM} MM},
  2021, pp. 1692--1701.

\bibitem{crmijcai20}
F.~Li, R.~Cong, H.~Bai, and Y.~He, ``Deep interleaved network for image
  super-resolution with asymmetric co-attention,'' in \emph{Proc. IJCAI}, 2020,
  pp. 534--543.

\bibitem{crmgcl2019tip}
C.~Guo, C.~Li, J.~Guo, R.~Cong, H.~Fu, and P.~Han, ``Hierarchical features
  driven residual learning for depth map super-resolution,'' \emph{IEEE Trans.
  Image Process.}, vol.~28, no.~5, pp. 2545--2557, 2019.

\bibitem{crmGCPANet}
Z.~Chen, Q.~Xu, R.~Cong, and Q.~Huang, ``Global context-aware progressive
  aggregation network for salient object detection,'' in \emph{AAAI}, 2020, pp.
  10\,599--10\,606.

\bibitem{intro-4wang2017central}
S.~Wang \emph{et~al.}, ``Central focused convolutional neural networks:
  Developing a data-driven model for lung nodule segmentation,'' \emph{Med.
  Imag. Anal.}, vol.~40, pp. 172--183, 2017.

\bibitem{DBLP:journals/tim/PramanikGBN20}
S.~Pramanik, S.~Ghosh, D.~Bhattacharjee, and M.~Nasipuri, ``Segmentation of
  breast-region in breast thermogram using arc-approximation and
  triangular-space search,'' \emph{IEEE Trans. Instrum. Meas.}, vol.~69, no.~7,
  pp. 4785--4795, 2020.

\bibitem{intro-5cherukuri2017learning}
V.~Cherukuri, P.~Ssenyonga, B.~C. Warf, A.~V. Kulkarni, V.~Monga, and S.~J.
  Schiff, ``Learning based segmentation of {CT} brain images: application to
  postoperative hydrocephalic scans,'' \emph{IEEE Trans. Biomed. Eng.},
  vol.~65, no.~8, pp. 1871--1884, 2017.

\bibitem{crmbrain}
Y.~Huang, F.~Zheng, R.~Cong, W.~Huang, M.~R. Scott, and L.~Shao, ``{MCMT-GAN:}
  multi-task coherent modality transferable {GAN} for {3D} brain image
  synthesis,'' \emph{{IEEE} Trans. Image Process.}, vol.~29, pp. 8187--8198,
  2020.

\bibitem{crmpolyp}
G.~Yue, W.~Han, B.~Jiang, T.~Zhou, R.~Cong, and T.~Wang, ``Boundary constraint
  network with cross layer feature integration for polyp segmentation,''
  \emph{IEEE J. Biomed. Health Inform.}, early access, doi:
  10.1109/JBHI.2022.3173948.

\bibitem{DBLP:journals/tim/DairiHS22}
A.~Dairi, F.~Harrou, and Y.~Sun, ``Deep generative learning-based {1-SVM}
  deteors for unsupervised {COVID-19} infection detection using blood tests,''
  \emph{IEEE Trans. Instrum. Meas.}, vol.~71, pp. 1--11, 2022.

\bibitem{DBLP:journals/tim/SharmaKMR21}
R.~R. Sharma, M.~Kumar, S.~Maheshwari, and K.~P. Ray, ``{EVDHM-ARIMA}-based
  time series forecasting model and its application for {COVID-19} cases,''
  \emph{IEEE Trans. Instrum. Meas.}, vol.~70, pp. 1--10, 2021.

\bibitem{maskrcnn}
K.~He, G.~Gkioxari, P.~Doll{\'{a}}r, and R.~B. Girshick, ``Mask {R-CNN},'' in
  \emph{Proc. ICCV}, 2017, pp. 2980--2988.

\bibitem{YOLACT}
D.~Bolya, C.~Zhou, F.~Xiao, and Y.~J. Lee, ``{YOLACT:} {R}eal-time instance
  segmentation,'' in \emph{Proc. ICCV}, 2019, pp. 9156--9165.

\bibitem{deeplab}
L.~Chen, Y.~Zhu, G.~Papandreou, F.~Schroff, and H.~Adam, ``Encoder-decoder with
  atrous separable convolution for semantic image segmentation,'' in
  \emph{Proc. ECCV}, 2018, pp. 833--851.

\bibitem{FCN}
J.~Long, E.~Shelhamer, and T.~Darrell, ``Fully convolutional networks for
  semantic segmentation,'' in \emph{Proc. CVPR}, 2015, pp. 3431--3440.

\bibitem{unet}
O.~Ronneberger, P.~Fischer, and T.~Brox, ``{U-Net}: Convolutional networks for
  biomedical image segmentation,'' in \emph{Proc. MICCAI}, 2015, pp. 234--241.

\bibitem{p1-gao2019res2net}
S.~Gao, M.-M. Cheng, K.~Zhao, X.-Y. Zhang, M.-H. Yang, and P.~H. Torr,
  ``{Res2Net}: A new multi-scale backbone architecture,'' \emph{IEEE Trans.
  Pattern Anal. Mach. Intell.}, vol.~43, no.~2, pp. 652--662, 2021.

\bibitem{p2-woo2018cbam}
S.~Woo, J.~Park, J.-Y. Lee, and I.~S. Kweon, ``{CBAM}: Convolutional block
  attention module,'' in \emph{Proc. ECCV}, 2018, pp. 3--19.

\bibitem{zhang2020dense}
Q.~Zhang \emph{et~al.}, ``Dense attention fluid network for salient object
  detection in optical remote sensing images,'' \emph{IEEE Trans. Image
  Process.}, vol.~30, pp. 1305--1317, 2021.

\bibitem{zhang2019Etnet}
Z.~Zhang, H.~Fu, H.~Dai, J.~Shen, Y.~Pang, and L.~Shao, ``{ET-Net}: A generic
  edge-attention guidance network for medical image segmentation,'' in
  \emph{Proc. MICCAI}, 2019, pp. 442--450.

\bibitem{l1-qin2019basnet}
X.~Qin, Z.~Zhang, C.~Huang, C.~Gao, M.~Dehghan, and M.~Jagersand, ``{BASNet}:
  Boundary-aware salient object detection,'' in \emph{Proc. CVPR}, 2019, pp.
  7479--7489.

\bibitem{l2-wei2020f3net}
J.~Wei, S.~Wang, and Q.~Huang, ``F$^3${Net}: Fusion, feedback and focus for
  salient object detection,'' in \emph{Proc. AAAI}, vol.~34, no.~07, 2020, pp.
  12\,321--12\,328.

\bibitem{segdata1}
``{COVID-19} {CT} segmentation dataset,'' \url{https://
  medicalsegmentation.com/COVID19/}, accessed April, 2020.

\bibitem{segdata2}
``{COVID-19} {CT} lung and infection segmentation dataset,''
  \url{https://zenodo.org/record/3757476}, accessed April 20, 2020.

\bibitem{2020Lung}
F.~Shan \emph{et~al.}, ``Lung infection quantification of {COVID-19} in {CT}
  images with deep learning,'' \emph{arXiv preprint arXiv:2003.04655}, 2020.

\bibitem{S}
D.-P. Fan, M.-M. Cheng, Y.~Liu, T.~Li, and A.~Borji, ``Structure-measure: A new
  way to evaluate foreground maps,'' in \emph{Proc. ICCV}, 2017, pp.
  4548--4557.

\bibitem{E}
D.-P. Fan, C.~Gong, Y.~Cao, B.~Ren, M.-M. Cheng, and A.~Borji,
  ``Enhanced-alignment measure for binary foreground map evaluation,'' in
  \emph{Proc. IJCAI}, 2018, pp. 4548--4557.

\bibitem{crm2019tcsvt}
R.~Cong, J.~Lei, H.~Fu, M.-M. Cheng, W.~Lin, and Q.~Huang, ``Review of visual
  saliency detection with comprehensive information,'' \emph{IEEE Trans.
  Circuits Syst. Video Technol.}, vol.~29, no.~10, pp. 2941--2959, 2019.

\bibitem{unet++}
Z.~Zhou, M.~M.~R. Siddiquee, N.~Tajbakhsh, and J.~Liang, ``{UNet++}: A nested
  {U-Net} architecture for medical image segmentation,'' in \emph{Proc.
  MICCAIW}, 2018, pp. 3--11.

\bibitem{attentionunet}
O.~Oktay \emph{et~al.}, ``Attention {U-Net}: Learning where to look for the
  pancreas,'' \emph{arXiv preprint arXiv:1804.03999}, 2018.

\bibitem{Cenet}
Z.~Gu \emph{et~al.}, ``{CE-Net}: Context encoder network for {2D} medical image
  segmentation,'' \emph{IEEE Trans. Med. Imag}, vol.~38, no.~10, pp.
  2281--2292, 2019.

\bibitem{Res2Net}
S.-H. Gao, M.-M. Cheng, K.~Zhao, X.-Y. Zhang, M.-H. Yang, and P.~Torr,
  ``{Res2Net}: A new multi-scale backbone architecture,'' \emph{IEEE Trans.
  Pattern Anal. Mach. Intell.}, vol.~43, no.~2, pp. 652--662, 2021.

\bibitem{imagenet}
J.~Deng, W.~Dong, R.~Socher, L.-J. Li, K.~Li, and L.~Fei-Fei, ``Imagenet: A
  large-scale hierarchical image database,'' in \emph{Proc. CVPR}, 2009, pp.
  248--255.

\bibitem{DBLP:conf/wacv/LaradjiRMLLKP0N21}
I.~H. Laradji \emph{et~al.}, ``A weakly supervised consistency-based learning
  method for {COVID-19} segmentation in {CT} images,'' in \emph{Proc. WACV},
  2021, pp. 2452--2461.

\bibitem{adam}
D.~Kingma and J.~Ba, ``Adam: A method for stochastic optimization,''
  \emph{arXiv preprint arXiv:1412.6980}, 2017.

\bibitem{dosovitskiy2020image}
A.~Dosovitskiy \emph{et~al.}, ``An image is worth 16x16 words: {T}ransformers
  for image recognition at scale,'' \emph{arXiv preprint arXiv:2010.11929},
  2020.

\bibitem{liu2021swin}
Z.~Liu \emph{et~al.}, ``Swin transformer: {H}ierarchical vision transformer
  using shifted windows,'' in \emph{Proc. ICCV}, 2021, pp. 10\,012--10\,022.

\end{thebibliography}
}

\end{document}